# Ultra-fast calorimetric measurements of the electronic heat capacity of graphene

Mohammed Ali Aamir[1†], John N. Moore[1†], Xiaobo Lu[1], Paul Seifert[1], Dirk Englund[2], Kin Chung Fong[3,4] and Dmitri K. Efetov[1*]

1. ICFO - Institut de Ciencies Fotoniques, The Barcelona Institute of Science and Technology, Castelldefels, Barcelona 08860, Spain
2. Department of Electrical Engineering and Computer Science, Massachusetts Institute of Technology, Cambridge, Massachusetts 02139, United States
3. Quantum Information Processing Group, Raytheon BBN Technologies, Cambridge, Massachusetts 02138, United States
4. Department of Physics, Harvard University, Cambridge, Massachusetts 02138, United States

*Email: dmitri.efetov@icfo.eu   †These authors contributed equally to this work.

**Heat capacity is an invaluable quantity in condensed matter physics, yet it has been so far experimentally inaccessible in two-dimensional (2D) van der Waals (vdW) materials, owing to their ultra-fast thermal relaxation times and the lack of suitable nano-scale thermometers. Here, we demonstrate a novel thermal relaxation calorimetry scheme that allows the first measurements of the electronic heat capacity of graphene $C_e$. It is enabled by the grouping of a radio-frequency Johnson noise thermometer, which can measure the electronic temperature $T_e$ with a measurement sensitivity of $\delta T_e \sim 20$ mK, and an ultra-fast photo-mixed optical heater, which can simultaneously modulate $T_e$ with a frequency of up to $\Omega = 0.2$ THz. This combination allows record sensitive and record fast measurements of the electronic heat capacity $C_e < 10^{-19}$ J/K, with an electronic thermal relaxation time $\tau_e < 10^{-13}$, representing orders of magnitude improvements as compared to previous state-of-the-art calorimeters. These features embody a breakthrough in heat capacity metrology of nano-scale and low-dimensional systems, and provide a new avenue for the investigation of their thermodynamic quantities.**

Heat capacity $C$ is a thermodynamic quantity which contains direct information of the energy ground state of a physical system. Measurements of $C$ were instrumental in laying the foundation of modern solid state physics as they gave crucial insight into the bosonic nature of lattice vibrations and the free electron gas in metals (*1*), and lead to the discovery of complex quantum phases in systems like heavy fermion compounds (*2*), frustrated spin glasses (*3*), high-T$_c$ superconductors (*4*) and of non-abelian topological states (*5*, *6*). Defined as $C = \partial E/\partial T$, where $E$ is the internal energy of the system and $T$ is its temperature, it can be measured by dynamical heat exchange of a given sample with a thermal bath. Here some of the most prominent techniques are AC (*7*) and relaxation (*8*) calorimetry, which simultaneously measure the systems thermal relaxation time $\tau$ and its thermal dissipation into the environment, which is defined by the thermal conductivity $G_{th}$. In the adiabatic limit $C$ is then obtained through the simple relation (*8*):

$$C = G_{th}\tau \qquad [1]$$

However, calorimetry of nano-scale and low-dimensional systems is notoriously challenging, since $C$ scales with volume and becomes much smaller than that of suitable thermometers. In addition, thermometry on such systems is often undermined by substantial heat leaks to the environment (*7*) and ultra-fast relaxation processes. Current state-of-the-art techniques for nano-scale systems have improved by many orders of magnitude in the last decades and use all electric radio-frequency (RF) techniques that allow measurements of $C \sim 10^{-19}$ J/K. Here the latest improvements were primarily enabled by pushing the time resolution to $\tau \sim 10^{-9}$ s (*6, 8–16*), however, these timescales are typically still not suitable for the smallest and thinnest nano-compounds from the class of two-dimensional vdW materials (*17–19*) and especially graphene (*20–22*), which have heat capacities and thermal relaxation times which are orders of magnitude lower and faster than what current experimental schemes can measure.

Here, we report on the development of a novel relaxation calorimetry scheme that allows record sensitive and ultra-fast measurements of the heat capacity of electrons $C_e$ of nano-scale objects down to values of $C_e \sim 5.9 \times 10^{-20}$ J/K and with electronic thermal relaxation times $\tau_e$ as fast as $\tau_e \sim 3 \times 10^{-13}$ s, which we apply to demonstrate the first measurement of $C_e$ of graphene. This is achieved by devising a combined ultra-fast optical heating and electronic thermometry readout scheme. It can measure the electronic temperature $T_e$ of nano-scale conductors with a measurement sensitivity of $\delta T_e \sim 20$ mK/Hz$^{1/2}$ (SI) and simultaneously modulate $T_e$ with a frequency of up to $\Omega = 0.2$ THz (Fig. 1A). The general concept of the measurement of $C_e$ is described in Fig. 1B, where a near-IR laser is focused onto the sample and selectively heats only the electrons in the sample raising its temperature by $\Delta T_e = T_e - T_L$ above the lattice temperature $T_L$, which the laser does not couple to (Fig. 1C). In the linear response regime of low heating power $P$, $\Delta T_e$ is defined by $\Delta T_e = P/G_{th}$, where $G_{th}$ is the electronic thermal conductivity which is typically given by electron-electron and electron-phonon scattering mechanisms (Fig. 1D). After the laser beam is turned off, $T_e$ exponentially cools off to $T_L$ with a thermal relaxation time which is defined by $\tau_e = G_{th}/C_e$ (Fig. 1E). Hence mutual measurements of $G_{th}$ and $\tau_e$ allow to directly probe $C_e$.

To demonstrate the measurement scheme we employ a two-terminal graphene device which is encapsulated by hexagonal boron nitride (hBN) and is locally gated by a bottom graphite layer, which allows to capacitively control its electron density $n$ by applying a gate voltage. It rests on an insulating sapphire (Al$_2$O$_3$) substrate, which is chosen to minimize parasitic capacitance and signal losses. To measure $T_e$ of the device we read out the thermal Johnson noise of its electrons, which conveniently has a white spectral power $P_{JN} = k_B T_e B$, where $k_B$ is Boltzmann's constant and $B$ is the frequency bandwidth of the measurement (*21, 22*). Our JN setup is comprised of an electrical radio frequency (RF) circuit which efficiently transmits the emitted $P_{JN}$ from the device through a 50Ω transmission line via an on-chip impedance matched LC network. $P_{JN}$ is then amplified with a series of low noise amplifiers and filters and integrated with a Schottky diode square-law detector, which linearly converts the noise power to a DC voltage $V_{JN}$ (Fig. 1A).

The transmission band is defined by the device resistance $R$ and the *LC* matching circuit and is identified around a resonance frequency $f_{res} \sim 107$ MHz with $B \sim 25$ MHz, which is marked by a sharp drop in the reflection parameter $S_{11}$ (Fig. 2A). However, both the $f_{res}$ and $B$ are strongly dependent on $n$, which significantly alters the device resistance $R(n)$, which has a characteristic peak shape at the charge neutral point where $n = 0$. The measured JN power spectral density (PSD) through this band is highly temperature dependent (Fig. 2B) and its amplitude is directly proportional to $k_B T_e$. It allows to calibrate the thermometer output $V_{JN}$ against the cryostat's built-in thermometer $T_L (= T_e)$ (Fig. 2C). Overall, we find that the JN thermometer has a sensitivity of $\delta T_e \sim 17$ mK for a measurement integration time $\tau_m = 1$s, at 15 K [see Methods and SI], which is obtained from a calibration with the Dicke radiometry formula, $\delta T_e = (T_e + T_{sys})/\sqrt{\tau_m B}$, which relates the standard deviation $\delta T_e$ with the system noise temperature $T_{sys}$, as shown in the inset of Fig. 2C. $\delta T_e$ can be further improved by increasing the impedance-matching bandwidth and cooling of the low-noise amplifier (*21*, *23*, *24*).

For the measurement of $\tau_e$ we use an optical setup in which two frequency-detuned near-IR lasers interfere, where the first laser has power $P_1$ and is tunable with a wave-length $\lambda_1 = 1548$-$1552$ nm and the second laser has power $P_2$ and a wave-length $\lambda_2 = 1550$ nm (Fig. 1A). This creates an amplitude-modulated beam with a total power $P$, which has a beating frequency $\Omega = 2\pi c(\lambda_1^{-1} - \lambda_2^{-1})$ that we can change with a precision of $\sim 0.2$ THz by tuning $\lambda_1$:

$$P = P_1 + P_2 + 2\sqrt{P_1 P_2} \sin \Omega t \qquad [2]$$

The laser beam is focused onto the sample through an optical objective inside the cryostat, which creates a collimated beam with a diffraction-limited spot of $\sim 2$ μm. The absorbed laser power in the graphene is normalized to only about 1.1% of the incident laser power, as is calculated by the optical transfer matrix method of the hBN encapsulated graphene device [see Methods].

In response to the heating by the laser beam $T_e$ oscillates with the same beating frequency $\Omega$, as is simulated in Fig. 2D using realistic parameters. These oscillations are however distorted sinusoids, whose amplitude becomes dampened as $\Omega$ grows, and with growing $\Omega$, the cooling half-cycles are dampened more than the heating half-cycles. This is due to the saturating power dependence of $\Delta T_e(P)$ (Fig. 1D), which becomes non-linear at higher $P$ because of a $G_{th}$ which is increasing with $T_e$. Hence $T_e$ cannot fully respond to changes in $P$ on the time scale of $\tau_e$, resulting in an increase of the time-averaged value $\langle T_e \rangle$ with $\Omega$, as is elucidated in Fig. 2E and F. In general, $\langle T_e \rangle$ has a Lorentzian dependence on $\Omega$ with a full width half maximum (FWHM) which is equal to $(\pi \tau_e)^{-1}$ for very weak heating power (see analytical derivation in SI):

$$\langle T_e(\Omega) \rangle = T_L + F_1(P_1, P_2) - P_1 P_2 \left[ F_2(P_1, P_2) + F_3(P_1, P_2) \frac{1/\tau_e^2}{(1/\tau_e^2 + \Omega^2)} \right], \qquad [3]$$

where $F_1$, $F_2$ and $F_3$ are positive functions of the decoupled $P_1$ and $P_2$, which depend on $G_{th}(T_L)$. With our setup we selectively measure only the $\Omega$ dependent part of $\langle T_e(\Omega) \rangle$ which is represented by the last term in Eq. 3, and which we call $|\Delta T_\Delta(\Omega)|$. It is measured by chopping both lasers and demodulating the thermometer readout signal at the chopping difference frequency $\Delta$ (see Fig. 1A and Methods). We can then directly extract $\tau_e$ from the resulting peak in $|\Delta T_\Delta(\Omega)|$ at $\Omega = 0$ THz (Fig. 1E).

We now apply the above presented techniques to directly measure $G_{th}$ and $\tau_e$ of the graphene device. Fig. 3A plots $G_{th}$ and Fig. 3B plots $\tau_e$ as a function of carrier density $n$ at bath temperatures $T_L = 15.5$, 60 and 100 K, and the insets plot both quantities as a function of $T_L$ at a fixed $n = -0.38 \times 10^{12}$ cm$^{-2}$. We model and fit $G_{th}$ and $\tau_e$ with established thermal mechanisms for graphene and find very good agreement, both in the density and temperature dependence. Past studies on graphene have identified two dominant mechanisms by which hot electrons cool in this range of temperatures: 1) $G_{WF}$ is the diffusive electronic heat transport to the thermally anchored electrodes and 2) $G_{ep}$ is the electron-acoustic phonon scattering (*22, 25, 26*).

The diffusive cooling power is given by $G_{WF} = \nabla(\kappa \nabla T_e)$, where the in-plane thermal conductivity of charge carriers $\kappa = L_0 T_e \sigma$ obeys the Wiedemann-Franz law, with $L_0$ being the Lorenz number and $\sigma$ the electronic conductivity (*21*). The electron-phonon cooling power is given by $G_{ep} = A\Sigma_{ep}(T_e^\delta - T_L^\delta)$, where $A$ is the device channel area, $\Sigma_{ep}$ is the electron-phonon coupling coefficient and $\delta$ is an exponent ranging from 3 to 4 that depends on disorder and carrier density of the device (*22, 25, 26*). Due to the differing functional dependences of $G_{WF}$ and $G_{ep}$ on $T_L$, there is a crossover temperature below which $G_{WF} > G_{ep}$. The dominance of diffusion cooling may be seen qualitatively at $T_L = 15.5$ K with the strong dip in $G_{th}$ that appears at charge neutrality, which is caused by the minimum in $\sigma$ that occurs there [SI]. As we raise $T_L$, the effect of electron-phonon cooling becomes observable; this leads to a less pronounced dip at charge neutrality due to the weaker density dependence (*22*) of $\Sigma_{ep} \sim \sqrt{n}$ relative to $\sigma \sim n$. The performed fits of these models with the data yield $\delta = 4.37$ and $\Sigma_{ep} = 6.9 \times 10^{-5}$ W m$^{-2}$K$^{-\delta}$ and a cross-over temperature of 126 K, which are consistent with earlier reports on graphene devices (*22, 26*). To fit the experimental value of $\tau_e$ we use the relation of $C_e/G_{th}$, in which $C_e$ is the theoretically calculated heat capacity and $G_{th}$ is the experimentally measured thermal conductance.

We finally combine the above measurements to extract the heat capacity of the graphene electrons through the relation $C_e = G_{th}\tau_e$, and normalize it to the device area to obtain the specific heat $c_e$. We show the so obtained $c_e$ values and the parameter free theoretical fits (see SI) as a function of $n$ for $T_L = 15.5, 60$ and $100$ K (Fig. 4A), and as a function of temperature $c_e$ vs $T_L$ (for $n = -0.38 \times 10^{12}$ cm$^{-2}$) (Fig. 4B). Notably, we find excellent qualitative and quantitative match of experiment and theory, where $c_e \propto \sqrt{n}$ scales in agreement to the expected proportionality of $c_e$ to the density of states (*21*), in addition to the expected linear temperature dependence $c_e \propto T_e$.

In conclusion, we compare the experimental performance of our calorimetry scheme with previous small-sample calorimeters in Fig. 4C. It allows measurement of the smallest specific heat values which were ever reported, with record low values of only $c_e \sim 270$ $k_B/\mu m^2$ (at charge neutrality and at base temperature), and with a record calorimetric sensitivity of only $\delta c_e \sim 36 \ k_B/\mu m^2$ (which can be extracted from the corresponding error bars). Moreover, our calorimeter allows the measurement of $c_e$ in a material with a record fast $\tau_e <$ 0.3ps. These characteristics represent a breakthrough in heat capacity metrology, and so open up the pathway for heat capacity measurements of two-dimensional materials, where it can be used to directly probe exotic ground states and quantum phase transitions. Our work is also interesting and timely because it relates to the currently burgeoning area of quantum thermodynamics (*27*) in which a major aspect is calorimetry of quantum objects (*28*), where heat transport is studied in the quantum regime.

1. N. W. Ashcroft, N. D. Mermin, *Solid State Physics* (1976).
2. R. A. Fisher, S. Kim, B. F. Woodfield, N. E. Phillips, L. Taillefer, K. Hasselbach, J. Flouquet, A. L. Giorgi, J. L. Smith, Specific heat of UPt3: Evidence for unconventional superconductivity. *Physical Review Letters*. **62**, 1411–1414 (1989).
3. A. P. Ramirez, A. Hayashi, R. J. Cava, R. Siddharthan, B. S. Shastry, Zero-point entropy in "spin ice." *Nature*. **399**, 333–335 (1999).
4. T. Park, M. B. Salamon, E. M. Choi, H. J. Kim, S. I. Lee, Evidence for Nodal Quasiparticles in the Nonmagnetic Superconductor YNi2B2C via Field-Angle-Dependent Heat Capacity. *Physical Review Letters*. **90**, 177001 (2003).
5. E. Sela, Y. Oreg, S. Plugge, N. Hartman, S. Lüscher, J. Folk, Detecting the Universal Fractional Entropy of Majorana Zero Modes. *Physical Review Letters*. **123**, 147702 (2019).
6. B. A. Schmidt, K. Bennaceur, S. Gaucher, G. Gervais, L. N. Pfeiffer, K. W. West, Specific heat and entropy of fractional quantum Hall states in the second Landau level. *Physical Review B*. **95**, 201306 (2017).
7. G. Ventura, M. Perfetti, *Thermal Properties of Solids at Room and Cryogenic Temperatures* (Springer Netherlands, Dordrecht, 2014; http://link.springer.com/10.1007/978-94-017-8969-1), *International Cryogenics Monograph Series*.
8. R. Bachmann, F. J. DiSalvo, T. H. Geballe, R. L. Greene, R. E. Howard, C. N. King, H. C. Kirsch, K. N. Lee, R. E. Schwall, H. U. Thomas, R. B. Zubeck, Heat capacity measurements on small samples at low temperatures. *Review of Scientific Instruments*. **43**, 205–214 (1972).
9. K. L. Viisanen, J. P. Pekola, Anomalous electronic heat capacity of copper nanowires at sub-Kelvin temperatures. *Physical Review B*. **97**, 115422 (2018).
10. O. Bourgeois, S. E. Skipetrov, F. Ong, J. Chaussy, Attojoule calorimetry of mesoscopic superconducting loops. *Physical Review Letters*. **94**, 057007 (2005).
11. S. Tagliati, V. M. Krasnov, A. Rydh, Differential membrane-based nanocalorimeter for high-resolution measurements of low-temperature specific heat. *Review of Scientific Instruments*. **83**, 055107 (2012).
12. S. G. Doettinger-Zech, M. Uhl, D. L. Sisson, A. Kapitulnik, Simple microcalorimeter for measuring microgram samples at low temperatures. *Review of Scientific Instruments*. **72**, 2398–2406 (2001).
13. A. Comberg, S. Ewert, W. Sander, Calorimeter for heat capacity measurements on quench-condensed thin metallic films in the range 0.6-4.4 K. *Cryogenics*. **18**, 79–81 (1978).
14. F. Fominaya, T. Fournier, P. Gandit, J. Chaussy, Nanocalorimeter for high resolution measurements of low temperature heat capacities of thin films and single crystals. *Review of Scientific Instruments*. **68**, 4191–4195 (1997).
15. R. L. Greene, C. N. King, R. B. Zubeck, J. J. Hauser, Specific Heat of Granular Aluminum Films. *Physical Review B*. **6**, 3297 (1972).
16. E. Pinsolle, A. Rousseau, C. Lupien, B. Reulet, Direct Measurement of the Electron Energy Relaxation Dynamics in Metallic Wires. *Physical Review Letters*. **116**, 236601 (2016).
17. Y. H. Wang, D. Hsieh, E. J. Sie, H. Steinberg, D. R. Gardner, Y. S. Lee, P. Jarillo-Herrero, N. Gedik, Measurement of Intrinsic Dirac Fermion Cooling on the Surface of the Topological Insulator Bi2Se3 Using Time-Resolved and Angle-Resolved Photoemission Spectroscopy. *Physical Review Letters*. **109**, 127401 (2012).

18. A. Sterzi, A. Crepaldi, F. Cilento, G. Manzoni, E. Frantzeskakis, M. Zacchigna, E. van Heumen, Y. K. Huang, M. S. Golden, F. Parmigiani, SmB6 electron-phonon coupling constant from time- and angle-resolved photoelectron spectroscopy. *Physical Review B*. **94**, 081111 (2016).
19. S. Roth, A. Crepaldi, M. Puppin, G. Gatti, D. Bugini, I. Grimaldi, T. R. Barrilot, C. A. Arrell, F. Frassetto, L. Poletto, M. Chergui, A. Marini, M. Grioni, Photocarrier-induced band-gap renormalization and ultrafast charge dynamics in black phosphorus. *2D Materials*. **6**, 031001 (2019).
20. M. M. Jadidi, R. J. Suess, C. Tan, X. Cai, K. Watanabe, T. Taniguchi, A. B. Sushkov, M. Mittendorff, J. Hone, H. D. Drew, M. S. Fuhrer, T. E. Murphy, Tunable Ultrafast Thermal Relaxation in Graphene Measured by Continuous-Wave Photomixing. *Physical Review Letters*. **117**, 257401 (2016).
21. K. C. Fong, K. C. Schwab, Ultrasensitive and Wide-Bandwidth Thermal Measurements of Graphene at Low Temperatures. *Physical Review X*. **2**, 031006 (2012).
22. K. C. Fong, E. E. Wollman, H. Ravi, W. Chen, A. a. Clerk, M. D. Shaw, H. G. Leduc, K. C. Schwab, Measurement of the Electronic Thermal Conductance Channels and Heat Capacity of Graphene at Low Temperature. *Physical Review X*. **3**, 041008 (2013).
23. S. Gasparinetti, K. L. Viisanen, O.-P. Saira, T. Faivre, M. Arzeo, M. Meschke, J. P. Pekola, Fast Electron Thermometry for Ultrasensitive Calorimetric Detection. *Physical Review Applied*. **3**, 014007 (2015).
24. J. Crossno, X. Liu, T. A. Ohki, P. Kim, K. C. Fong, Development of high frequency and wide bandwidth Johnson noise thermometry. *Applied Physics Letters*. **106**, 023121 (2015).
25. M. W. Graham, S. F. Shi, D. C. Ralph, J. Park, P. L. McEuen, Photocurrent measurements of supercollision cooling in graphene. *Nature Physics*. **9**, 103–108 (2013).
26. A. C. Betz, F. Vialla, D. Brunel, C. Voisin, M. Picher, A. Cavanna, A. Madouri, G. Fève, J. M. Berroir, B. Plaçais, E. Pallecchi, Hot electron cooling by acoustic phonons in graphene. *Physical Review Letters*. **109**, 056805 (2012).
27. J. P. Pekola, Towards quantum thermodynamics in electronic circuits. *Nature Physics*. **11**, 118–123 (2015).
28. B. Karimi, F. Brange, P. Samuelsson, J. P. Pekola, Reaching the ultimate energy resolution of a quantum detector. *Nature Communications*. **11**, 367 (2020).

Acknowledgements

We thank E. Dias, M. M. Jadidi, J. Garcia de Abajo and A. Principi for useful discussions. D. K. E. acknowledges the support from Ministry of Economy and Competitiveness of Spain through the "Severo Ochoa" program for Centres of Excellence in R&D (SE5-0522), Fundació Privada Cellex, Fundació Privada Mir-Puig, the Generalitat de Catalunya through the CERCA program, the H2020 Programme under grant agreement n° 820378, Project: 2D·SIPC and the La Caixa Foundation; J. N. M acknowledges support from Marie Skłodowska-Curie grant agreement 754510; P.S. acknowledges support from the Alexander von Humboldt Foundation and the German Federal Ministry for Education and Research through the Feodor-Lynen program; and K.C.F. acknowledges support from the US Army Research Office under Cooperative Agreement number W911NF-17-1-0574.

Author contributions

D. K. E. and D. E. conceived the experiments; M. A. A., J. N. M and D. K. E. designed the experiments; M. A. A. fabricated the devices; M. A. A., J. N. M., D. K. E. and K. C. F. conducted the experiments and analyzed the data; X. L., P. S. and K. C. F. provided technical support. M. A. A., J. N. M. and D. K. E. wrote the paper.

**Supplementary Information** is available for this paper.

**Correspondence and requests for materials** should be addressed to D. K. E.

**Competing financial and non-Financial interests:**
The authors declare no competing financial and non-financial interests.

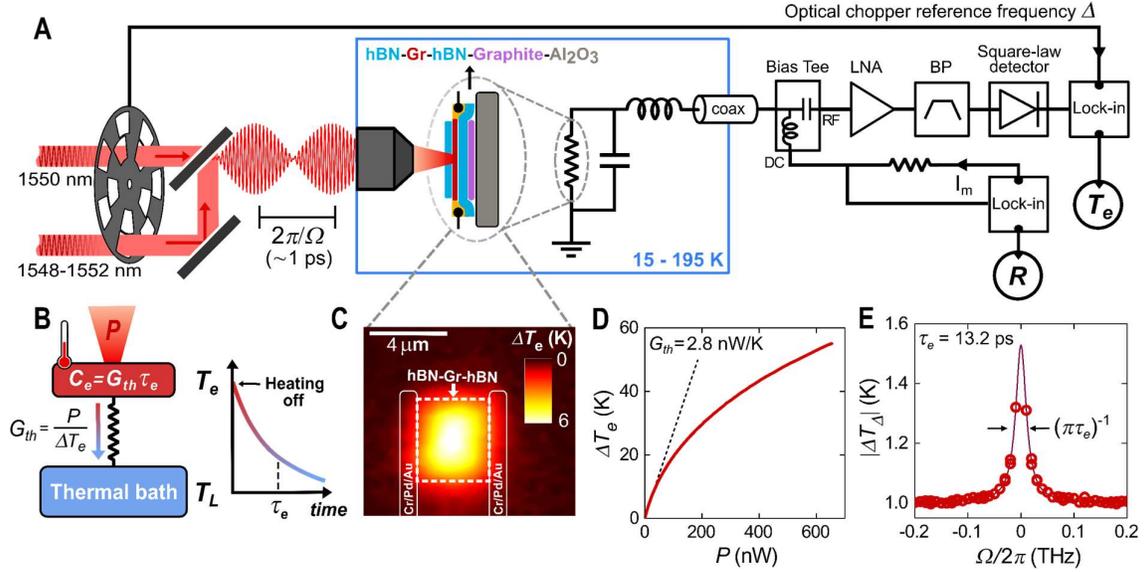

**Fig. 1. Experimental setup and its measurables.** (**A**) Schematic of the experimental setup. (**B**) Schematic of graphene's charge carriers (red body, at temperature $T_e$, with electronic heat capacity $C_e = G_{th}\tau_e$) in thermal contact (conductance $G_{th}$) with the thermal bath (blue body, at temperature $T_L$) provided by the graphene lattice and metal electrodes. $\tau_e$ is the characteristic temperature relaxation time when the heating is turned off. (**C**) Spatial map of electron temperature change $\Delta T_e$ when a CW monochromatic laser spot of power $P$ = 26 nW scans the device at charge neutrality and lattice temperature $T_L$ = 15.5 K. (**D**) $\Delta T_e$ as a function of $P$ incident at the device center. The slope in the quasi-linear-response regime near zero $P$ is the inverse of $G_{th} = P/\Delta T_e$. (**E**) Magnitude of $\Delta T_\Delta$, the non-linear component of time-averaged temperature change, as a function of beating frequency $\Omega$; solid line is a fit of Lorentzian function of width $(\pi\tau_e)^{-1}$.

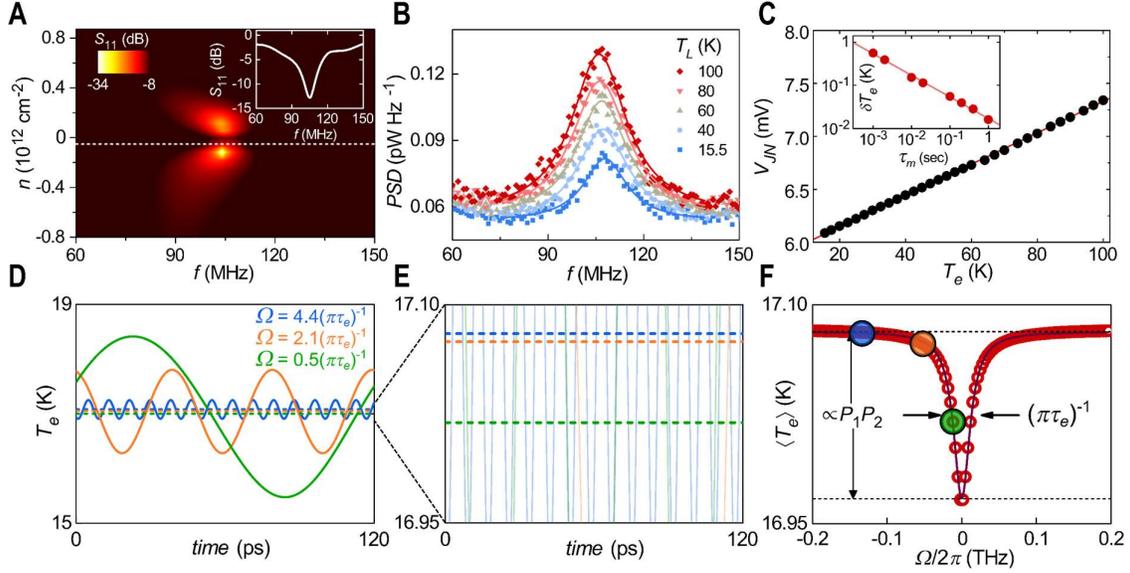

**Fig. 2. Thermometer operation and $\tau_e$ measurement principle.** **(A)** Reflection coefficient $S_{11}$ measured at the output of the impedance-matching LC circuit as a function of frequency and carrier density; dashed line indicates a line cut at $n = -0.051 \times 10^{12}$ cm$^{-2}$ that is plotted in the inset. **(B)** Amplified power spectral density of noise emitted from the device as a function of device temperature; solid lines are Lorentzian fits. **(C)** DC voltage output of the Schottky square-law detector as a function of the device temperature with a red line as a guide to eye; all data correspond to carrier density $n = -0.051 \times 10^{12}$ cm$^{-2}$. Inset: standard deviation in $T_e$ vs measurement integration time $\tau_m$ obeying Dicke radiometry formula indicated by the red line fit. **(D)** Simulated response of $T_e$ in time under oscillating heating power at three different frequencies $\Omega$; dashed lines indicate their time-averaged values $\langle T_e \rangle$. **(E)** Magnified view of (D) showing that $\langle T_e \rangle$ varies with $\Omega$. **(F)** Red circles: $\langle T_e \rangle$ vs. $\Omega$ obtained from simulated oscillations as in (D, E); the three highlighted points correspond to the time-domain data in (D, E); solid line is a Lorentzian fit to simulated data.

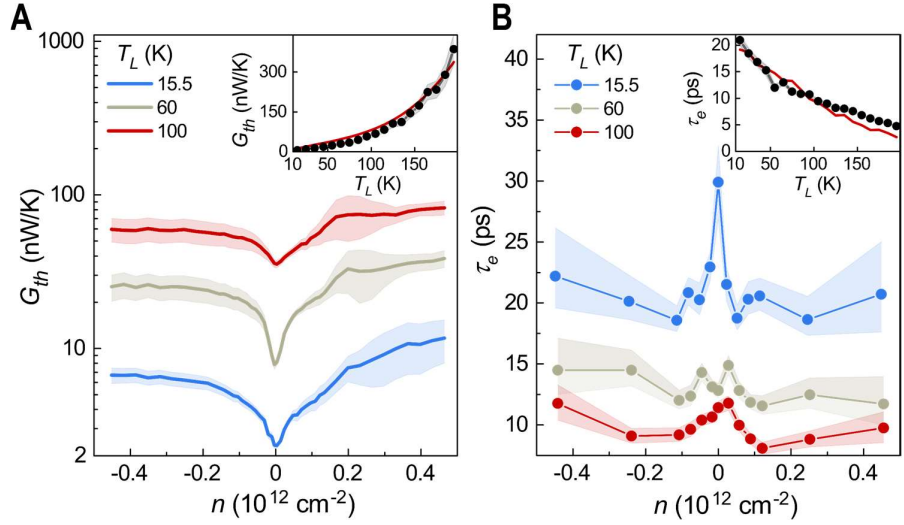

**Fig. 3. Measured $G_{th}$ and $\tau_e$.** (A) $G_{th}$ as a function of carrier density $n$ at three values of lattice temperature $T_L$. Inset: $G_{th}$ as a function of $T_e$ at $n = -0.38 \times 10^{12}$ cm$^{-2}$, where the red curve is a fit based on a model incorporating carrier diffusion cooling and electron-phonon cooling. Uncertainty in $G_{th}$ is dominated by uncertainty in contact resistance of the device. (B) $\tau_e$ as a function of $n$ at three values of $T_L$. Inset: $\tau_e$ as a function of $T_e$ at $n = -0.38 \times 10^{12}$ cm$^{-2}$, where the red curve is the expected $\tau_e = C_e/G_{th}$ calculated from theoretical $C_e$ and measured $G_{th}$ from the inset of (A). Uncertainty in $\tau_e$ is dominated by the standard deviation in measured data.

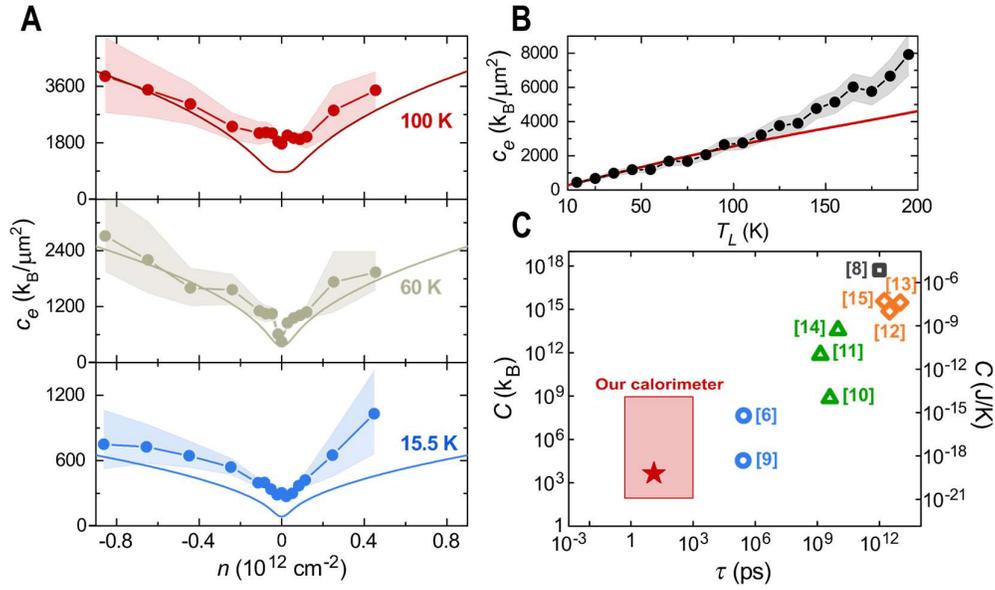

**Fig. 4. Electronic heat capacity of graphene.** (**A**) Electronic heat capacity per unit area, $c_e$, as a function of $n$ for three values of $T_L$ compared with theoretical calculations depicted as smooth curves. (**B**) $c_e$ as a function of $T_e$ at $n = -0.38 \times 10^{12}$ cm$^{-2}$ compared with theoretical calculation (red curve). Uncertainty in $c_e$ has propagated from uncertainties in measured $G_{th}$ and $\tau_e$. (**C**) A performance comparison of the calorimeter in this work (red star) with others designed for small samples from Ref. (6, 8–15), where the lowest measured heat capacity $C$ and $\tau$ in each calorimeter are plotted; the shaded region shows the range of applicability of our calorimeter.

## Methods:

Materials and fabrication

The devices are fabricated using a van der Waals assembly technique of crystals micro-mechanically exfoliated on a $Si^{++}/SiO_2$ (285 nm) surface. A thin hexagonal boron nitride (hBN) flake is picked up by a propylene carbonate (PC) film supported by polydimethyl siloxane (PDMS) at 100 ºC. The hBN flake is then used to pick up a monolayer graphene flake; a second hBN flake is next picked up in order to encapsulate the graphene by hBN. Finally, a graphite flake is picked up to serve as a local bottom gate electrode for tuning the graphene carrier density. The second hBN flake thickness is chosen by optical contrast to be ~ 30 nm and further confirmed with atomic force microscopy. The graphite gate is a few layers of graphene, with thickness > 1 nm. The final heterostructure is then deposited onto an insulating sapphire ($Al_2O_3$) substrate chosen to minimize parasitic capacitance and microwave losses. The device is then etched into a transfer length method (TLM) geometry having multiple graphene channels of varying length using CHF3/O2 plasma in the ratio 4:40 sccm flow. The graphene channel is edge-contacted by metallization of Cr/Pd/Au (2/15/50 nm). The edge contacts contribute an additional resistance separate from that of graphene, which we measure in order to subtract the contribution of the contacts to the Johnson-noise read-out [see details in SI].

Johnson-noise thermometry

The sapphire substrate of the device and the LC network are connected to a printed circuit board a PFT dielectric chosen for its high thermal conductance and coplanar waveguides as 50 Ω transmission lines. A surface mount inductor and capacitor made up the LC network, and had values 390 nH and 4.5 pF respectively. These values are chosen such that the RLC circuit formed by the graphene channel resistance and the LC network has a resonance frequency $f_{res} = (1/2\pi)\sqrt{1/LC - 1/(RC)^2} \approx 107$ MHz while also matching to the 50 Ω cable impedance. By the Dicke radiometry formula, $\delta T_e/(T_e + T_{sys}) = 1/\sqrt{\tau_m B}$, the standard deviation in measured temperature, $\delta T_e$ depends on the captured frequency bandwidth $B$ and measurement integration time $\tau_m$. $T_{sys}$ is the system noise temperature, measured to be 70 K, from which the thermometer sensitivity is determined at 15 K [see inset of Fig. 2C and SI]. Since $B = f_{res}/Q$, where $Q$ is the Q-factor of the LC network, a high $f_{res}$ is used to achieve a larger $B$ and reduce $\delta T_e$. The coaxial cable is made of stainless steel and connects to 1st- and 2nd-stage low-noise amplifiers (Caltech CITLF3 and Miteq AU-1263 respectively) through a bias tee (Mini-Circuits ZFBT-4R2GW+). The bias tee allows simultaneous measurement of RF noise and graphene 2-terminal resistance throughout all experiments. The diode square-law power detector (Fairview Microwave SMD0102) was read out with a multimeter (Keithley 2700) during thermometer calibration and using lock-in detection using the reference of the optical chopper modulating the incident heating laser light.

Optical transfer matrix calculation

The electric field amplitude within the device is determined for normally incident 1550 nm linearly polarized light by calculating the transfer matrix, which is obtained from Maxwell's equations, and accounts for the wave interference and propagation of light through all the layers of the stacked heterostructure. The transfer matrix is a function of the complex refractive indices of the layers and their thicknesses. The graphene layer in this calculation absorbs 1.1% of light incident on the device, independent of carrier density used in this experiment.

Extraction of $\tau_e$

The Lorentzian term in Eq. 3 is much smaller than $T_e$; therefore, to measure it the signal-to-noise ratio must be improved by chopping the two laser beams at different frequencies (here 270 Hz and 162 Hz). This ensures that only the two nonlinear terms in $\langle T_e \rangle$, which are scaled by the product $P_1 P_2$, are modulated at the chopping difference frequency $\Delta = 108$ Hz because the optical interference giving rise to these nonlinear terms only occur when both lasers are present. A lock-in amplifier is used to demodulate the detected temperature signal at $\Delta$, giving the measured amplitude $|\Delta T_\Delta|$. This amplitude is equal to the amplitude of the two nonlinear terms in Eq. 3, which comprise the Lorentzian and a comparably sized $\Omega$-independent offset, both negative in sign. See SI for an analytical proof that this Lorentzian will manifest in any material having a sublinear temperature response to heating power.

We observe experimentally and in simulation that as $\Delta T_e / T_L$ increases at a given $T_L$, there is a monotonic decrease in the value of $\tau_e$ directly extracted from the FWHM of the peak in $T_e$ [SI]. This again reflects the fact that the cooling power increases with $\Delta T_e$. The dependence of as-extracted $\tau_e$ on $\Delta T_e / T_L$ also provides a way to improve the determination of $\tau$ in the ground state. Using the numerical simulation of this dependence at each experimental condition, we interpolate from the measurement at finite $\Delta T_e / T_L$ to obtain a value of $\tau_e$ corresponding to the ground-state condition $\Delta T_e / T_L = 0$. We report here the interpolated ground-state $\tau_e$, which is always larger than the as-extracted $\tau_e$ by less than a factor of 2.

# Supporting Information: Measurement of the electronic heat capacity of graphene

Mohammed Ali Aamir[1†], John N. Moore[1†], Xiaobo Lu[1], Paul Seifert[1], Dirk Englund[2], Kin Chung Fong[3,4], Dmitri K. Efetov[1*]

5. ICFO - Institut de Ciencies Fotoniques, The Barcelona Institute of Science and Technology, Castelldefels, Barcelona 08860, Spain
6. Department of Electrical Engineering and Computer Science, Massachusetts Institute of Technology, Cambridge, Massachusetts 02139, United States
7. Quantum Information Processing Group, Raytheon BBN Technologies, Cambridge, Massachusetts 02138, United States
8. Department of Physics, Harvard University, Cambridge, Massachusetts 02138, United States

*Email: dmitri.efetov@icfo.eu   †These authors contributed equally to this work.

Contents

---

## 1. Johnson Noise Thermometer

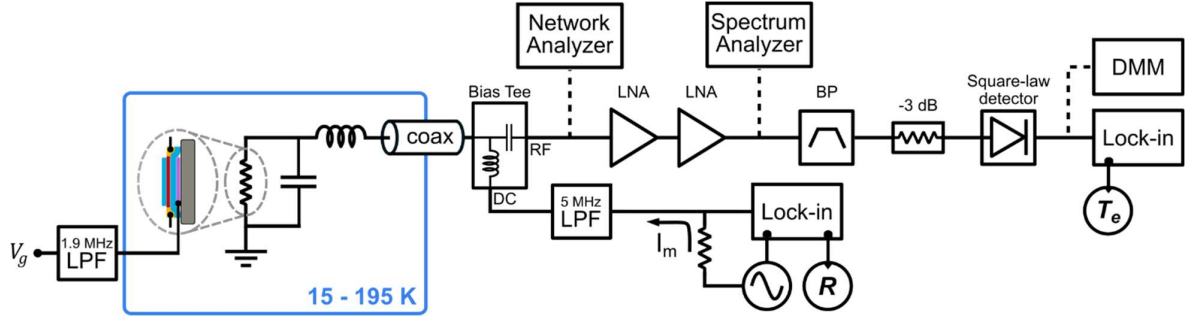

**Fig. S1** Schematic of the Johnson noise thermometer circuit.

Figure S1 is an elaborated version of the RF circuit presented in Fig. 1A. The circuit is designed to amplify voltage noise emitted from the resistive element, which is the device under test. This device is wire bonded and fixed to a copper printed circuit board by silver paste, which sits inside an AttoDry800 cryostat cooled down to a base temperature of 15 K. A local 4-W heater is used to raise the PCB temperature to a chosen temperature up to 195 K in this experiment. To capture the voltage noise emitted by the device, it is connected inside an LC impedance matching network (nominally $L = 390$ nH and $C = 4.5$ pF) which is matched to the 50-Ω input of the first-stage amplifier (Caltech LF3) over a bandwidth of ~25 MHz. The impedance matching is evaluated by measuring the parameter $S_{11}$ (related to the reflection coefficient Γ as $S_{11} = 20\log_{10}|\Gamma|$) using a network analyzer (Agilent 8753ES) attached to the RF port of the bias tee by a 2-metre long coaxial cable. Experimentally measured $S_{11}$ is plotted in Fig. 2A as a function of carrier density and frequency at a temperature of 15.5 K. Over this range of carrier density, $S_{11}$ has a dip depth ≤ -8 dB, corresponding to $1 - \Gamma^2 \geq 0.84$. The inset of this figure is a line cut of $S_{11}$ vs frequency at $n = -0.051 \times 10^{12}$ cm$^{-2}$ (white dashed line). The measured $S_{11}$ agrees well with a calculation of $S_{11}$, shown in Fig. S2, that uses the measured value of device resistance, the nominal value of $L$, and a value of $C$ chosen as 5.8 pF. When $C$ used in the calculation is ~5.8 pF rather than the nominal 4.5 pF the resonance frequency obtained at the global $S_{11}$ minimum is 104 MHz, which matches the resonance frequency observed. This indicates that the measurement circuit contains a parasitic capacitance of about 5.8 pF – 4.5 pF = 1.3 pF.

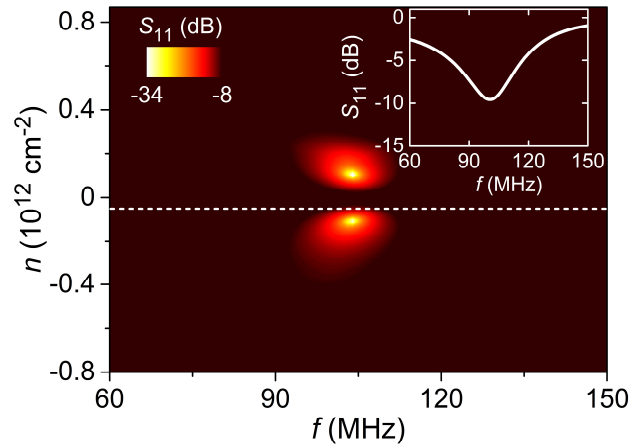

**Fig. S2** Calculated $S_{11}$ as a function of frequency and carrier density. Insets: Line cut at $n = -0.051 \times 10^{12}$ cm$^{-2}$ to directly compare with the measured values in the inset of Fig. 2A.

The power spectral density (PSD) (shown in Fig. 2B) is measured using a spectrum analyzer (Agilent N9320A) connected to the output of the second-stage amplifier while the bias tee and first-stage amplifier are directly connected. The peak in PSD of noise emitted by the device occurs at 107 MHz, which is a slightly higher frequency than the dip occurring in $S_{11}$. The difference may be due to a relative increase in resonance frequency of the LC network caused by large cable capacitance in the case of the $S_{11}$ measurement. The PSD may be described as the sum of a Johnson noise component, originating with the sample, and a temperature-independent system noise component $P_{sys}(\Gamma)$ contributed by the amplifiers and external sources:

$$\text{PSD}(v) = 4k_B T_e \Delta v g(1 - \Gamma(v)^2) + P_{sys}(\Gamma)$$
$$= 4k_B \Delta v g(1 - \Gamma(v)^2)(T_e + T_{sys}(\Gamma)).$$

Here, $v$ is the frequency, $\Delta v$ is the resolution bandwidth of spectrum analyzer, and $g$ is the total gain of the amplifiers. $T_{sys}$ is the system noise temperature, which serves to describe the electronic temperature at which Johnson noise and system noise are equal. $T_{sys}$ is measured by collecting the PSD at several temperatures and extracting the x-intercept of PSD vs. $T_e$; this value of $T_e$ at which PSD = 0 is $T_{sys}$. $T_{sys}$ is plotted as a function of carrier density and frequency in Fig. S3. $T_{sys}$ shows minima at the same carrier densities at which $S_{11}$ displays dips because the impedance matching at these conditions give the Johnson noise component maximum strength.

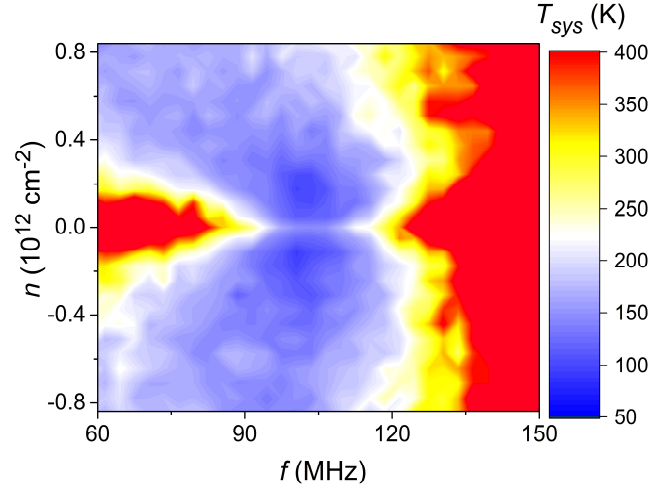

**Fig. S3** System noise temperature as a function of frequency and carrier density as extracted from linear fits of *PSD* vs temperature.

In the experiment, the temperature readout of the thermometer is performed using a diode square-law detector (Fairview Microwave SMD0102) instead of the spectrum analyzer. This detector is preceded by a band-pass filter (41-140 MHz) for the purpose of cutting out excess noise that does not originate in the device. A 3 dB attenuator is used to limit the signal to the detector to ensure that it responds linearly. The detector integrates the PSD in the spectral region of the LC resonance and outputs a voltage proportional to the integrated power. This voltage therefore scales linearly with the Johnson noise of the device and may be calibrated against the Cernox thermometer to give a readout of $T_e$. This calibration is described in the next section.

After calibrating the thermometer, its sensitivity is evaluated by measuring the standard deviation of its DC readout $\delta T_e$ using a digital multimeter. Figure S4 shows the distribution of $T_e$ measurements obtained using two different integration times $\tau_m$ of the thermometer readout at 15 K, which form two of the datapoints in the inset of Fig. 2C showing $\delta T_e$ as a function of the integration time $\tau_m$. The trend in Fig. 2C inset matches the prediction of the Dicke radiometry formula (*1*)

$$\delta T_e = \frac{T_e + T_{sys}}{\sqrt{\tau_m B}}$$

where $B$ is the effective bandwidth established by the band-pass of the LC impedance matching network before the diode detector, which is 25 MHz. A fit to this formula gives a system noise temperature for this thermometer configuration of 70 K at the charge neutrality point, corresponding to a sensitivity $\sqrt{S_{T_e}} = \sqrt{\tau_m \delta T_e^2} = (T_e + T_{sys})/\sqrt{B} = 17$ mK/Hz$^{1/2}$.

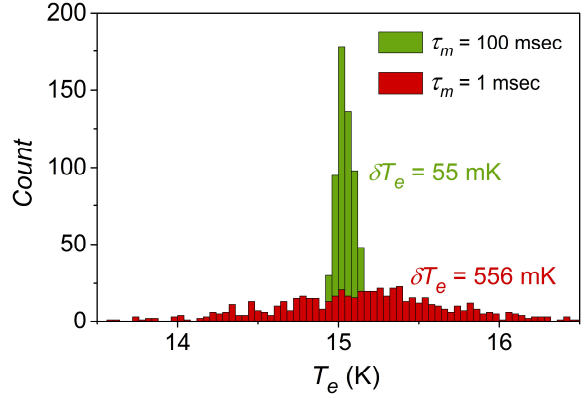

**Fig. S4** Histogram of electron temperature read-outs for two different measurement integration times $\tau_m$ using a digital multimeter at charge neutrality and a sample temperature of 15 K. $\delta T_e$ is the standard deviation found which obeys the Dicke radiometry formula.

2. Calibration of the Johnson Noise Thermometer

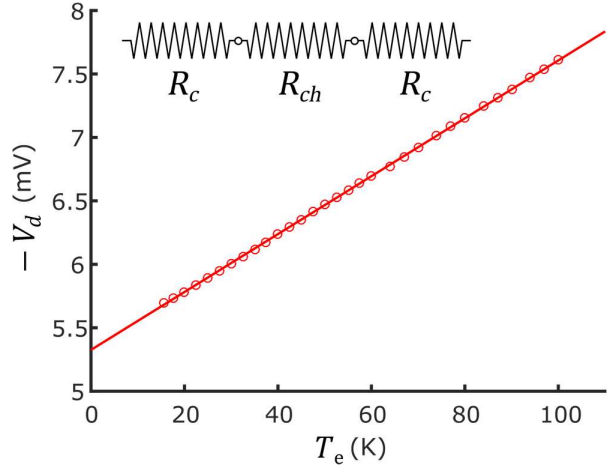

**Fig. S5** Diode power detector readout vs. electronic temperature $T_e$ of the device (same as $T_L$) when $n = -3.75 \times 10^{11}$ cm$^{-2}$ Inset: Three-resistor model of graphene channel and two contacts.

We model the resistance of our graphene device as the sum of three series components (Fig. S5 inset); these are the channel resistance $R_{ch}$ and two identical contact resistances $R_c$ created at the two graphene-electrode junctions (see figure inset). The Johnson noise voltage spectral density of elements in series combines additively; therefore, the total thermal voltage noise across our device in a bandwidth $\Delta f$ is

$$\langle V^2 \rangle = \langle V^2 \rangle_{ch} + \langle V^2 \rangle_c = 4k_B(T_{ch}R_{ch} + 2T_c R_c)\Delta f$$

where $T_{ch}$ and $T_c$ are the spatial average temperatures of the channel and contact regions. The total Johnson noise power emitted by the device in this bandwidth is, in units of watts

$$S_P \Delta f = \frac{\langle V^2 \rangle}{R}$$

where $R = R_{ch} + 2R_c$. Over the selected bandwidth the diode square-law detector receives an amplified power input of $G(S_P \Delta f) + P_{sys}$, where $G$ is the average power gain of the thermometer in that bandwidth. The diode detects this input power and converts it to a proportional output voltage

$$V_d = cG(S_P \Delta f) + cP_{sys} = 4k_B \Delta f \left( T_{ch} \frac{R_{ch}}{R} + 2T_c \frac{R_c}{R} \right) Gc + cP_{sys}$$

where $c$ is the diode's power-to-voltage conversion ratio, which is negative.

In a calibration of our noise thermometer, the graphene channel and contacts are both held in thermal equilibrium with the Cernox thermometer at temperature $T = T_{ch} = T_c$. Thus, in this case

$$V_d = 4k_B \Delta f GcT + cP_{sys}$$

By measuring $V_d$ against the Cernox thermometer readout, we obtain the slope of this curve (Fig. S5)

$$\left. \frac{dV_d}{dT} \right|_{meas} = 4k_B \Delta f Gc$$

Note that this slope changes as a function of carrier density because for this discussion $G$ contains the impedance-dependent coupling efficiency between the device and amplifier. For the case plotted in Fig. S5, the carrier density is $n = -3.75 \times 10^{11}$ cm$^{-2}$ and the magnitude of the local slope at 15.5 K is 21.3 µV/K.

$T_{ch}$ is the quantity we seek to measure for the purpose of obtaining $G_{th}$ of intrinsic graphene. However, during our experiment $T_{ch} \neq T_c$ because only the channel region is heated by the laser, while the contacts remain thermally well anchored to the thick gold electrodes which are effectively at the lattice temperature $T_L$. Therefore, to a 1$^{st}$-order approximation $T_c = T_L$, and we may use the relation

$$V_d = 4k_B \Delta f \left( T_{ch} \frac{R_{ch}}{R} + 2T_L \frac{R_c}{R} \right) Gc + cP_{sys}$$

giving

$$\frac{dV_d}{dT_{ch}} = \left. \frac{dV_d}{dT} \right|_{meas} \frac{R_{ch}}{R}$$

$$\Delta T_{ch} = \Delta V_d \left| \frac{dV_d}{dT} \right|_{meas}^{-1} \frac{R}{R_{ch}}$$

When the carrier density is $n = -3.75 \times 10^{11}$ cm$^{-2}$ and $T_L = 15.5$ K and the modulation of the diode voltage under the laser illumination has a magnitude $\Delta V_d = 6.36$ µV, this approximation gives an increase in channel temperature of $\Delta T_{ch} = 0.67$ K.

However, we do not use this method of approximation to calculate the results reported here. Rather, we improve on this 1$^{st}$-order approximation by considering that the local temperature of the contact region next to the electrodes may increase slightly above $T_L$ due to the fact that this region has some nonzero spatial extent away from the electrodes. Under the laser heating there exists a correlation between $T_c$ and $T_{ch}$, so we treat $T_c$ as a function dependent on $T_{ch}$. Then, our 2$^{nd}$-order approximation is

$$V_d = cG4k_B \Delta f \left( T_{ch} \frac{R_{ch}}{R} + 2T_c(T_{ch}) \frac{R_c}{R} \right) + cP_{sys}$$

giving

$$\frac{dV_d}{dT_{ch}} = \frac{dV_d}{dT}\bigg|_{meas} \left(\frac{R_{ch}}{R} + 2\frac{R_c}{R}\frac{dT_c}{dT_{ch}}\right)$$

We cannot directly measure the quantity $dT_c/dT_{ch}$, but we may arrive at a rough estimate by simulating the electronic temperature's spatial distribution in the graphene.

$$\frac{dT_c}{dT_{ch}} \approx \frac{\Delta T_c}{\Delta T_{ch}}\bigg|_{sim}$$

Here we estimate the derivative as the calculated ratio of average temperature increase in the contact regions $\Delta T_c$ to the average temperature increase in the channel region $\Delta T_{ch}$ under illumination matching the size and intensity used in the experiment. An example of one of the calculated temperature spatial distributions is shown in Fig. S6. This is calculated by numerically solving the heat equation in steady state:

$$P(x,y) = -\nabla \cdot [\kappa_{WF}(x,y)\nabla T_e(x,y)]$$

where $P$ is the laser heating power with a Gaussian profile, and $\kappa_{WF}$ is the in-plane thermal conductance calculated from the Wiedemann-Franz Law. Because of the different resistivities in the channel and in the contact region, $\kappa_{WF}$ will differ in these two regions. $\kappa_{WF} = T_e(x,y)\mathcal{L}_0 L/(WR_{ch})$ and $\kappa_{WF} = T_e(x,y)\mathcal{L}_0 l_c/(WR_c)$ inside the channel and contact regions respectively. $\mathcal{L}_0$ is the Lorenz ratio, $L$ and $W$ are the dimensions of the graphene, and $l_c$ is the length of the contact region. Our device has non-invasive 1D edge contacts which should create a region of contact resistance inside the graphene with a length less than the length of doping diffusion measured in 2D invasive contacts (~0.4 μm) (2). Therefore, we make this length as short as our grid spacing will allow (0.07 μm) and verify that further reducing this length produces no significant change in our result.

We simulate the diffusion of heat with the Wiedemann-Franz thermal conductance and neglect cooling to the lattice by phonon scattering because the phonon cooling strength is unknown to us prior to measuring $G_{th}$. Including a phonon cooling contribution can be regarded as a 3rd-order correction because it would reduce $dV_d/dT_{ch}$ by an amount smaller than the correction made between the 1st-order and 2nd-order models.

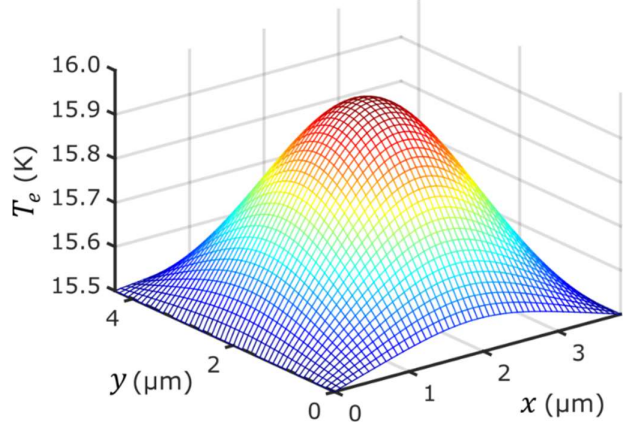

**Fig. S6** Spatial distribution of $T_e$ calculated in the graphene device using measured experimental parameters when $n = -3.75 \times 10^{11}$ cm$^{-2}$, $T_L = 15.5$ K. Grid spacing is ~0.7 μm. There is a segment of one grid length at the ends of the channel in which resistivity is increased by the presence of a contact resistance.

The final equation we use to calculate the results reported here is

$$\Delta T_{ch} = \Delta V_d \left|\frac{dV_d}{dT}\bigg|_{meas}\right|^{-1} \left(R_{ch} + 2R_c\frac{\Delta T_c}{\Delta T_{ch}}\bigg|_{sim}\right)^{-1} R$$

Under the same conditions as above ($n = -3.75 \times 10^{11}$ cm$^{-2}$, $T_L = 15.5$ K, $\Delta V_d = 6.36$ μV), $\Delta T_{ch} = 0.59$ K by this approximation.

Figure S7 plots the values of $\frac{dV_d}{dT_{ch}}$ obtained from our 1st- and 2nd-order approximations as a function of carrier density at 15.5 K.

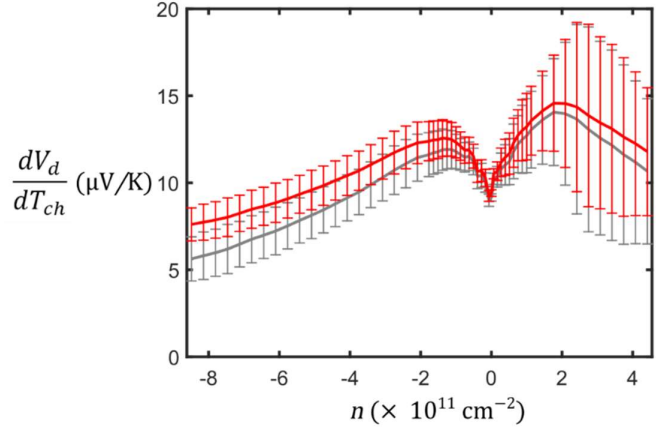

**Fig. S7** The calibration factor that converts the magnitude of measured diode detector voltage to a change in $T_{ch}$. The grey and red traces respectively have been obtained using the 1st- and 2nd-order approximations discussed in the text.

3. Optical Setup

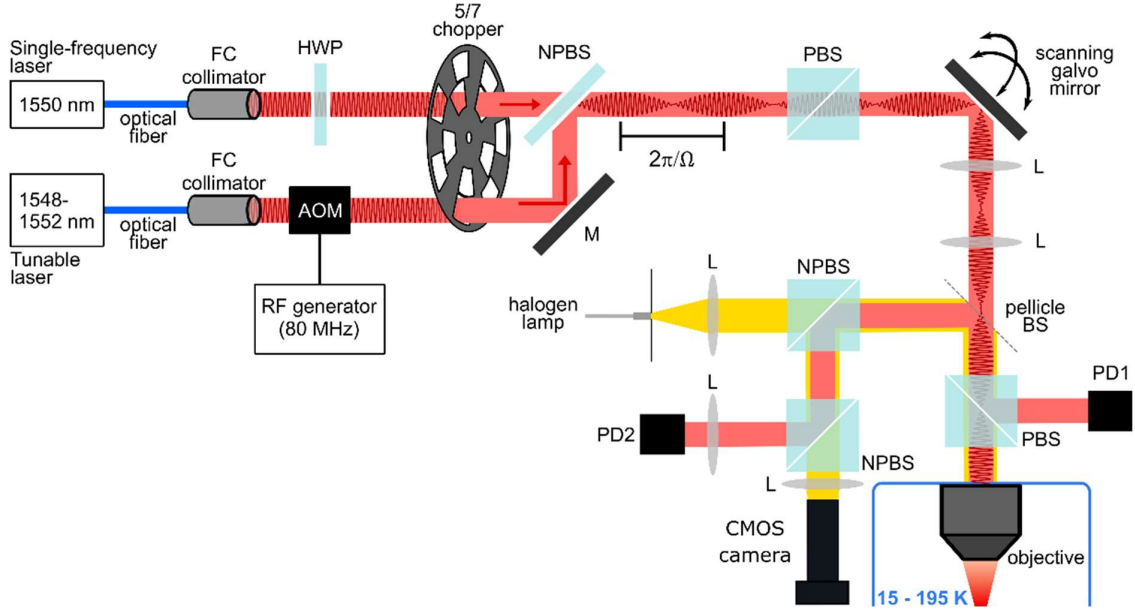

**Fig. S8** The optical setup used to heat the graphene device, spatially map its thermal response and perform optical alignment and focusing of the microscope.

Figure S8 shows the optical setup used to heat the graphene device. A single-frequency laser (Thorlabs SFL1550P) operating at 1550 nm and a tunable laser (Thorlabs TLX1), which we tune between 1548-1552 nm, interfere in this setup to create a beating in the resulting intensity at the difference frequency of the two lasers. The light from the fiber-coupled lasers is collimated and coupled to free space by two fiber-coupled collimating lenses. The two laser beams are aligned collinearly at the first non-polarizing beam splitter (NPBS) and then linearly polarized using a polarizing beam splitter (PBS). Absorption polarizers are not used because these are strongly achromatic. Before an experiment, a half-wave plate (HWP) is used to manually set the intensity of the single-frequency laser beam transmitted through the PBS. An acousto-optic modulator (AOM) (Gooch & Housego), powered by an RF generator, can control the fraction of light that transmits through it. During the measurement of $\tau_e$, the tunable laser intensity that arrives at the device is stabilized using the AOM in a PID loop. The PID receives feedback from the thermometer signal measured with a lock-in on the frequency at which the tunable laser is chopped. No additional steps are taken to stabilize the intensity of the single-frequency laser.

The two-axis scanning galvo mirror forms a 4-$f$ scanning galvo microscope with two spherical lenses and a 50X objective lens (Olympus LMPLN50XIR). This microscope has a scan area of 60 x 60 µm². This is used to perform scanning microscopy of the reflection intensity and the thermal response of the graphene. The reflected laser intensity is confocally detected with a photodiode (PD2). The focused laser spot size was measured using the reflected intensity detected while scanning a lithographically defined metal edge feature across the laser spot. To manipulate the device's position, it is placed on piezoelectric positioners (AttoCube). The device and piezoelectric positioners sit in an AttoDry800 cryostat. The device surface is also confocally imaged using white light from a halogen lamp and a CMOS camera in the visible spectrum. This aids in alignment and focusing of the device in the microscope.

The polarization ratio of the linearly polarized light is reduced as a result of reflecting off of the scanning galvo mirrors and other mirrors not shown here. A second PBS placed before the objective lens restores the lost linear polarization of the light prior to its arrival at the sample. Two PBSs are used in total so that fluctuation in the polarization of a laser source will be translated into a fluctuation in intensity before arriving at the second PBS. This way, the intensity ratio of the two paths of the second PBS remains fixed, and PD1 is a good monitor of the intensity at the device. The laser intensity is monitored using PD1 during measurements of $G_{th}$. To avoid effects of refraction through the second PBS, all experiments are performed while the scanning galvo mirrors are set to their origin and the laser is normally incident on this PBS.

4. Device Characteristics

We approximate the contact resistance at the terminals of the device by performing a transfer length measurement (TLM). The graphene channel used in the main experiments (length 3.84 µm) is the longest channel in the microscope image in Fig. S9A. Figure S9B shows TLM plots at four carrier densities measured at 15 K. The average combined resistance of both contacts $2R_c$ is plotted in Fig. S9C as a function of carrier density (red curve). The measured difference between the two-terminal resistance $R$ of the device and $2R_c$ is the channel resistance $R_{ch}$. $2R_c$ contains non-physical features near charge neutrality which are due to slight differences in quality from one channel to another. We therefore generate a modeled $2R_c$ as a Lorentzian peak at charge neutrality offset in $y$ by the minimum measured $2R_c$ values at high electron density and high hole density. The offset resistance is higher for negative doping than for positive doping because the contacts introduce $n$ doping at the graphene edge which results in a $p$-$n$ junction at the contacts when the channel is negatively doped by the gate. We use the modeled $2R_c$ in the rest of our data analysis.

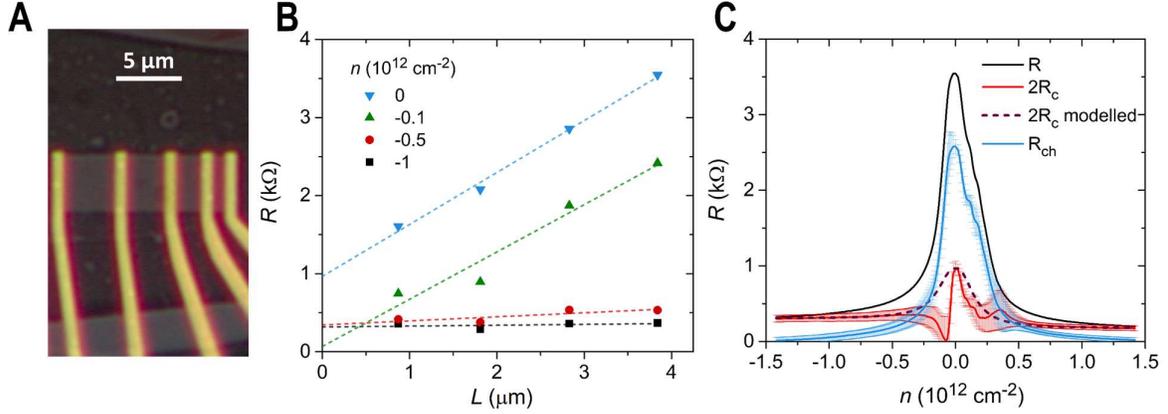

**Fig. S9 (A)** Optical image of four graphene channels contacted by Cr/Pd/Au electrodes via 1D edge contacts. Channel lengths are 0.87, 1.81, 2.83 and 3.84 μm. We use the longest channel in the main experiment. **(B)** The TLM, i.e. two-terminal resistance vs channel length. **(C)** Resistance of the longest channel. Two-terminal resistance $R$; combined contact resistance $2R_c$; channel resistance $R_{ch}$; and $2R_c$ modeled as a Lorentzian peak offset by minimum measured $2R_c$.

We approximate the density of electron and hole puddles at charge neutrality by analyzing the carrier-density width of the Dirac resistance peak. Figure S10A shows this peak measured in the two-terminal resistance of the device at 15.5 K at the beginning of the main experiment (three thermal cycles after the TLM). The two-terminal conductance, which is the inverse of the plotted resistance, is plotted as a function of carrier density in log scale in Fig. S10B. An extrapolation of the linear part of the conductance achieves the minimum measured value of conductance at a carrier density $n^*$. This $n^*$ is the magnitude of the spatial fluctuations in carrier density that persists throughout the channel at charge neutrality due to disorder (*3*).

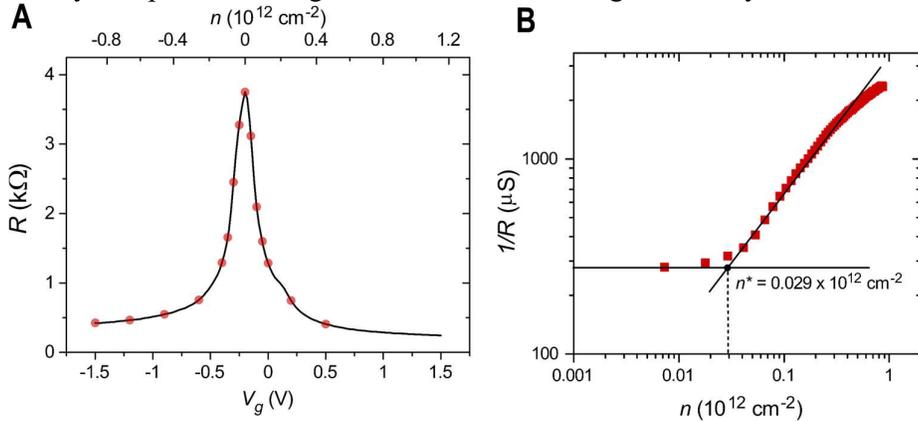

**Fig. S10 (A)** Two-terminal resistance vs carrier density of the main device measured at 15.5 K. Red markers indicate carrier densities at which we measure heat capacity. **(B)** Two-terminal conductance vs carrier density plotted using data at positive carrier densities in (A).

## 5. Fitting $G_{th}$ vs $T_e$

We fit the measured data plotted in the inset of Fig. 3A ($G_{th}$ vs $T_e$) using a nonlinear regression (MATLAB's *nlinfit*) with a modeling function that we now describe. Inputs to the modeling function are test values of two fitting parameters $\Sigma$ and $\delta$, and the values of $T_e$ at which we performed the measurements. The modeling function outputs values of $G_{th}$ evaluated for the input values of $T_e$. These $G_{th}$ values are calculated by simulating the steady-state spatial distribution of $\Delta T_e(x,y) = T_e(x,y) - T_L$ under a weak Gaussian heating profile $P(x,y)$. $G_{th}$ is evaluated as the ratio of the spatial averages of these two quantities $\Delta T_{e,avg}/P_{avg}$. Here, we choose the heating power to be small enough that $\Delta T_e$ responds linearly. We simulate $T_e(x, y)$ under the effects of both heat diffusion and phonon cooling terms according to the heat equation
$$P(x, y) = -\nabla \cdot [\kappa_{WF}(x, y)\nabla T_e(x, y)] + A\Sigma(T_e(x, y; t)^\delta - T_L^\delta),$$
where $A$ is the graphene area. This simulation also takes into account the resistance of the contact regions at the two ends of the channel, as discussed in the above "Calibration of the Johnson Noise Thermometer".

We performed this 2-parameter fitting not only for the dataset in the inset of Fig. 3A measured at $n = -0.38 \times 10^{12}$ cm$^{-2}$, but also for $G_{th}$ vs $T_e$ dataset measured at charge neutrality plotted in Fig. S11. At charge neutrality (at $n = -0.38 \times 10^{12}$ cm$^{-2}$), the best-fit electron-phonon coupling constant $\Sigma$ is $1.4 \pm 0.1 \times 10^{-4}$ ($6.9 \pm 0.9 \times 10^{-5}$) W m$^{-2}$ K$^{-\delta}$, and the best-fit cooling exponent $\delta$ is $4.2 \pm 0.2$ ($4.37 \pm 0.02$).

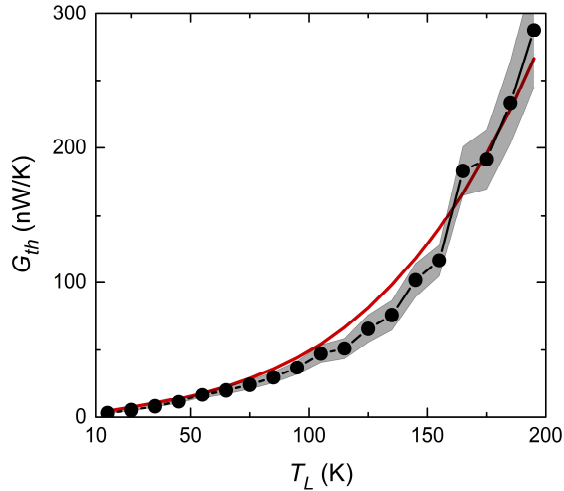

**Fig. S11** Thermal conductance vs electronic temperature measured at charge neutrality. Measurement error derives from uncertainty in contact resistance. Red curve is a best fit to the described diffusion model.

6. <u>Derivation of Lorentzian Peak in $\langle T_e(\Omega)\rangle$</u>

In the following, we derive Eq. 2 that appears in the main text, i.e.,

$$\langle T_e(\Omega)\rangle = T_L + F_1(P_1, P_2) - P_1 P_2 \left[F_2(P_1, P_2) + F_3(P_1, P_2) \frac{1/\tau_e^2}{\left(\frac{1}{\tau_e^2} + \Omega^2\right)}\right].$$

We make three assumptions: (1) The heating power is weak enough that $\Delta T_e \ll T_L$, where $\Delta T_e = T_e - T_L$; (2) $\tau_e$ is constant with temperature over the range $T_l$ to $T_L + \Delta T_e$; (3) $G_{th} = k_1 T_e^n + k_2 T_e^m$. Assumption (3) accounts for the possibility of multiple competing cooling mechanisms and implies further generalization to an arbitrary number of cooling mechanisms. Using the dynamic form of the heat equation

$$C_e \frac{dT_e}{dt} = P - Q$$

we wish to solve for $T_e(t)$ under the sinusoidal heating power $P(t)$ in Eq. 1. $C_e$ and $Q$ are the electronic heat capacity and cooling power.

$$G_{th}\tau_e \frac{dT_e}{dt} = P_1 + P_2 + 2\sqrt{P_1 P_2} \sin \Omega t - \int G_{th} dT_e$$

$$(k_1 T_e{}^n + k_2 T_e{}^m)\tau_e \frac{dT_e}{dt} = P_1 + P_2 + 2\sqrt{P_1 P_2}\sin\Omega t - \frac{k_1}{n+1}T_e{}^{n+1} - \frac{k_2}{m+1}T_e{}^{m+1} + c(T_L)$$

Substituting $T_e = \Delta T_e + T_L$,

$$\left[a_1\left(x^n + nx^{n-1}T_L + \cdots + \binom{n}{2}T_L^{n-2}x^2 + nT_L^{n-1}x + T_L^n\right)\right.$$
$$\left. + a_2\left(x^m + mx^{m-1}T_L + \cdots + \binom{m}{2}T_L^{m-2}x^2 + mT_L^{m-1}x + T_L^m\right)\right]\frac{dx}{dt}$$
$$= P_{dc} + \mathcal{A}\sin\Omega t$$
$$- b_1\left(x^{n+1} + (n+1)x^n T_L + \cdots + \binom{n+1}{2}T_L^{n-1}x^2 + (n+1)T_L^n x + T_L^{n+1}\right)$$
$$- b_2\left(x^{m+1} + (m+1)x^m T_L + \cdots + \binom{m+1}{2}T_L^{m-1}x^2 + (m+1)T_L^m x\right.$$
$$\left. + T_L^{m+1}\right) + c(T_L)$$

Here $c(T_L)$ is a constant of integration. We have used the substitutions $\tau_e k_i \equiv a_i$, $\frac{k_i}{n+1} \equiv b_i$, $P_1 + P_2 \equiv P_{dc}$, $2\sqrt{P_1 P_2} \equiv \mathcal{A}$, and $\Delta T_e \equiv x$.

By assumption (1), $x \ll T_L$; and we may expand $x$ as an ordered series of approximation—here we expand up to 3rd order:
$$x(t) = x^{(1)} + x^{(2)} + x^{(3)},$$
where $x^{(1)}$ and its time derivative have a linear response to $P(t)$, and higher orders respond linearly to higher powers of $P(t)$. Specifically, $\{x^{(N)}, \dot{x}^{(N)}\} \propto P^N$. Consistent with our 3rd-order approximation, in Eq. V we neglect terms in $x^{(N_1)}\dot{x}^{(N_2)}$ for $N_1 + N_2 > 3$. Eq. V then reduces to

$$\left[a_1\left(\binom{n}{2}T_L^{n-2}x^2 + nT_L^{n-1}x + T_L^n\right) + a_2\left(\binom{m}{2}T_L^{m-2}x^2 + mT_L^{m-1}x + T_L^m\right)\right]\frac{dx}{dt}$$
$$= P + \mathcal{A}\sin\Omega t$$
$$- b_1\left(\binom{n+1}{3}T_L^{n-2}x^3 + \binom{n+1}{2}T_L^{n-1}x^2 + (n+1)T_L^n x + T_L^{n+1}\right)$$
$$- b_2\left(\binom{m+1}{3}T_L^{m-2}x^3 + \binom{m+1}{2}T_L^{m-1}x^2 + (m+1)T_L^m x + T_L^{m+1}\right)$$
$$+ c(T_L)$$

Expanding the above in orders of 1, 2 and 3 yields the following system of three equations:

$$\frac{dx^{(1)}}{dt} + \gamma x^{(1)} = \beta + \alpha \sin\Omega t$$

$$\frac{dx^{(2)}}{dt} + \gamma x^{(2)} = -px^{(1)}\frac{dx^{(1)}}{dt} - q x^{(1)2}$$

$$\frac{dx^{(3)}}{dt} + \gamma x^{(3)} = -p\left(x^{(1)}\frac{dx^{(2)}}{dt} + x^{(2)}\frac{dx^{(1)}}{dt}\right) - rx^{(1)2}\frac{dx^{(1)}}{dt} - 2qx^{(1)}x^{(2)} - ux^{(1)3},$$

where $\gamma \equiv 1/\tau_e$, $\beta \equiv \frac{P_{dc} + c(T_L) - (b_1 T_L^{n+1} + b_2 T_L^{m+1})}{a_1 T_L^n + a_2 T_L^m}$, $\alpha \equiv \frac{\mathcal{A}}{a_1 T_L^n + a_2 T_L^m}$, $p \equiv \frac{a_1 n T_L^{n-1} + a_2 m T_L^{m-1}}{a_1 T_L^n + a_2 T_L^m}$, $q \equiv \frac{b_1\binom{n+1}{2}T_L^{n-1} + b_2\binom{m+1}{2}T_L^{m-1}}{a_1 T_L^n + a_2 T_L^m}$, $r \equiv \frac{a_1\binom{n}{2}T_L^{n-2} + a_2\binom{m}{2}T_L^{m-2}}{a_1 T_L^n + a_2 T_L^m}$, and $u \equiv \frac{b_1\binom{n+1}{3}T_L^{n-2} + b_2\binom{m+1}{3}T_L^{m-2}}{a_1 T_L^n + a_2 T_L^m}$.

After solving these equations, we obtain the time-independent components of $x^{(1)}$, $x^{(2)}$ and $x^{(3)}$ as

$$x_{DC}^{(1)} = \frac{\beta}{\gamma} = \frac{A}{B}$$

$$x_{DC}^{(2)} = -\frac{q}{\gamma}\left(\left(\frac{\beta}{\gamma}\right)^2 + \frac{\alpha^2}{2(\gamma^2 + \Omega^2)}\right) = -\frac{C}{B^3}\left(A^2 + \frac{2P_1P_2}{\tau_e^2\left(\frac{1}{\tau_e^2} + \Omega^2\right)}\right)$$

$$x_{DC}^{(3)} = \beta\left(\beta^2\frac{2q^2 - u\gamma}{\gamma^5} + \alpha^2 q\frac{2q\gamma + p\Omega^2}{\gamma^2(\gamma^2 + \Omega^2)^2} - \alpha^2\frac{3u\gamma - 2q^2}{2\gamma^3(\gamma^2 + \Omega^2)}\right)$$

$$= \frac{A}{B^3}\left[\frac{A^2}{B}\left(\frac{C^2}{B} - \frac{D}{3}\right) - \frac{4P_1P_2}{\tau_e^2\left(\frac{1}{\tau_e^2} + \Omega^2\right)}\left(\frac{1}{2}\frac{D}{B} - \frac{3}{4}\frac{C^2}{B^2}\right)\right],$$

where we define $A \equiv P_1 + P_2 + c(T_L) - \left(\frac{k_1}{n+1}T_L^{n+1} + \frac{k_2}{m+1}T_L^{m+1}\right)$, $B \equiv k_1 T_L^n + k_2 T_L^m$, $C \equiv k_1 n T_L^{n-1} + k_2 m T_L^{m-1}$, and $D \equiv k_1 n(n-1)(n-2)T_L^{n-2} + k_2 m(m-1)(m-2)T_L^{m-2}$. Therefore, the time-independent component of the increase in $T_e$ above $T_L$ has a Lorentzian dependence on $\Omega$ with FWHM equal to $(\pi\tau_e)^{-1}$:

$$\Delta T_{DC} = \frac{A}{B}\left[1 + \frac{A}{B^2}\left(\frac{A}{B}\left(\frac{C^2}{B} - \frac{D}{3}\right) - C\right)\right] - \frac{2}{B^3}\frac{1}{\tau_e^2}\frac{P_1P_2}{\left(\frac{1}{\tau_e^2} + \Omega^2\right)}\left(C + \frac{A}{B}\left(D - \frac{3}{2}\frac{C^2}{B}\right)\right).$$

Here, the 1st-order, 2nd-order and 3rd-order terms are colored in green, red and blue respectively.

We will now only consider the case that $\{n, m\} \geq 0$, which describes all the physical cooling mechanisms we are aware of. Then, $\{A, B, C, D\} \geq 0$, and $\Delta T_{DC}$ as a function of $\Omega$ may be visualized as in Fig. S12. Taking the value of $G_{th}$ at $T_L$, our analysis shows that 1st-, 2nd- and 3rd-order terms have sizes linear in $\frac{P_1+P_2}{G_{th}}$, $\left(\frac{P_1+P_2}{G_{th}}\right)^2 \frac{1}{T_L}$ and $\left(\frac{P_1+P_2}{G_{th}}\right)^3 \frac{1}{T_L^2}$ respectively. It follows from Assumption (1) that the size of these terms diminishes as their order increases. The result depicted in Fig. S12 is a Lorentzian dip that rides on top of an offset. This Lorentzian appears if either $n$ or $m$ is greater than zero.

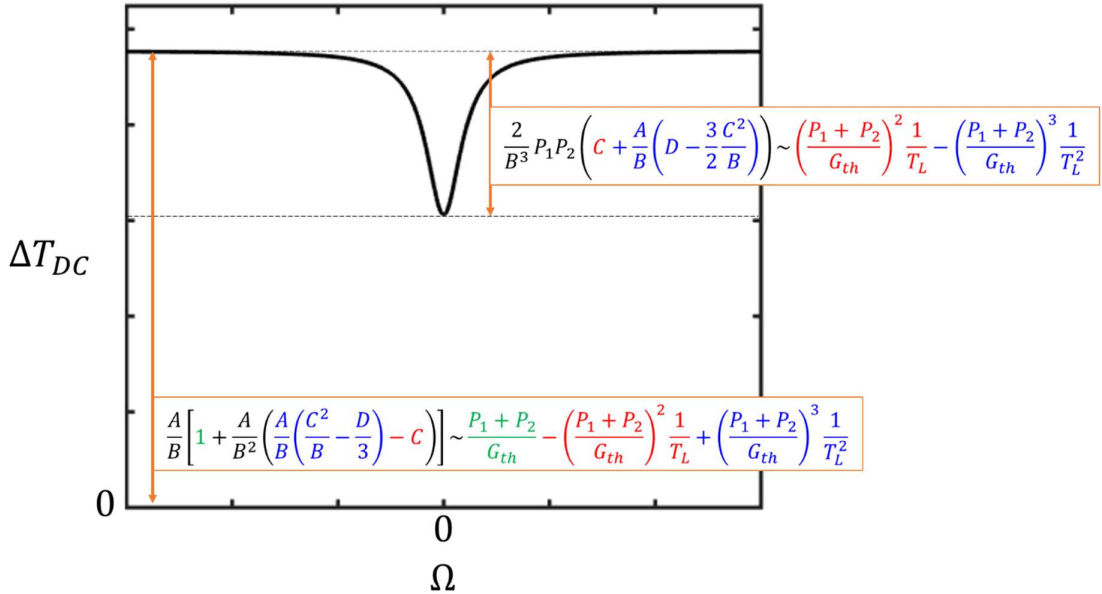

*Fig. S12 The analytically calculated time-average increase in electron temperature $\Delta T_{DC}$ vs $\Omega$ under AC heating power. The green, red and blue terms are respectively 1st-, 2nd- and 3rd-order components of $\Delta T_{DC}$, and have decreasing magnitude.*

At this point of analysis it is evident that the double-chopping experiment, which we conduct, only detects the signal component linear in $P_1 P_2$, will not be sensitive to the 1st-order term. It will only detect the components of the higher-order terms that are linear in $P_1 P_2$; therefore, we expect the experiment to detect a Lorentzian having an amplitude and an offset of the same sign. Indeed, this is what the experiment detects as $\Delta T_\Delta$ in Fig. 1E.

This point becomes more clear when we explicitly express $\Delta T_{DC}$ in terms of $P_1$ and $P_2$:

$$\Delta T_{DC} = \eta_1(P_1 + P_2) + \eta_2(P_1^2 + P_2^2) + \epsilon(P_1^3 + P_2^3) + \eta_3$$
$$- P_1 P_2 \left[ -(3\epsilon(P_1 + P_2) + 2\eta_2) + (\eta_4(P_1 + P_2) + \eta_5) \frac{1/\tau_e^2}{\left(\frac{1}{\tau_e^2} + \Omega^2\right)} \right],$$

where $\eta_1 \equiv \frac{1}{B} - \frac{2\xi C}{B^3} + 3\xi^2 \epsilon > 0$, $\eta_2 \equiv -\frac{C}{B^3} + 3\xi\epsilon < 0$, $\eta_3 \equiv \xi\left[\frac{1}{B} - \frac{\xi C}{B^3} + \xi^2 \epsilon\right] < 0$, $\eta_4 \equiv \frac{2}{B^4}\left(D - \frac{3}{2}\frac{C^2}{B}\right) < 0$, $\eta_5 \equiv \frac{2}{B^3}\left(C + \frac{\xi}{B}\left(D - \frac{3}{2}\frac{C^2}{B}\right)\right) > 0$, $\epsilon = \frac{1}{B^4}\left(\frac{C^2}{B} - \frac{D}{3}\right) > 0$, $\xi = c(T_L) - \left(\frac{k_1}{n+1}T_L^{n+1} + \frac{k_2}{m+1}T_L^{m+1}\right)$.

Then we may write the result in terms of two components, one of which is linear in the quantity $P_1 P_2$, and the other which is not:

$$\Delta T_{DC} = F_1(P_1, P_2) - P_1 P_2 \left[ F_2(P_1, P_2) + F_3(P_1, P_2) \frac{1/\tau_e^2}{\left(\frac{1}{\tau_e^2} + \Omega^2\right)} \right].$$

Here, $F_1$, $F_2$ and $F_3$ are all functions in which $P_1$ and $P_2$ are decoupled. Analysis also shows that these functions are all positive.

## 7. Normalizing Measured Thermal Relaxation Time

In the limit when an infinitesimal heating power is used for the measurement of $\tau_e$, $\tau_e$ may be extracted from the full-width at half maximum (FWHM) of the Lorentzian peaks, e.g. Fig. 1E, as described in the main text. Then the measured quantity $(\pi \cdot \text{FWHM})^{-1}$ is identical to the characteristic exponential relaxation time of $T_e$ after being perturbed from equilibrium. However, when using a finite heating power as we do in the experiment, the value obtained from the FWHM is less than $\tau_e$. We will call this quantity $\tau_e^*$.

$$\tau_e^* = (\pi \cdot \text{FWHM})^{-1} < \tau_e$$

We came to understand this fact first from numerical simulations of the experiment. We solve the following equation to calculate how $T_e$ changes in time under the oscillating heating power of the lasers:

$$C_e(x, y; t) \frac{dT_e(x, y; t)}{dt} = -A\Sigma\left(T_e(x, y; t)^\delta - T_L^\delta\right) + \nabla \cdot [\kappa_{WF}(x, y; t)\nabla T_e(x, y; t)] + P(x, y; t),$$

where $P = P_1(x, y) + P_2(x, y) + 2\sqrt{P_1(x, y)P_2(x, y)}\sin\Omega t$ is the laser heating power having a Gaussian profile, $\kappa_{WF} = T_e(x, y; t)\mathcal{L}_0 L/(WR_{ch})$ is the in-plane thermal conductance given by the Wiedemann-Franz Law, $A$ is the graphene area, and $\Sigma$ and $\delta$ are the electron-phonon cooling power coefficients. $C_e$ is the electronic heat capacity which we here estimate to be linear with changes in $T_e$ under the laser heating. At each point in time, we obtain a spatial distribution of $T_e$ resembling Fig. S6. The spatial average of $T_e$, which we will call $T_{avg}$ oscillates in time at frequency $\Omega$ as described in the main text. We then average over several of the oscillations of $T_{avg}$ to obtain the effective non-oscillating component that would be read out by a thermometer (Fig. S13). We call this value $T_{DC}$. Next, we plot the value of $T_{DC}$ as a function of $\Omega/2\pi$ and fit this to a Lorentzian function to extract the value of $\tau_e^*$ (Fig. S14). Then we repeat this calculation of $\tau^*$ for several different values of heating power $P$. Finally, we use these results to generate a plot of $\tau_e^*/\tau_e$ vs $\Delta T_e/T_L$, where $\Delta T_e$ is here taken as $T_{DC} - T_L$ near the base of the Lorentzian peak (Fig. S15).

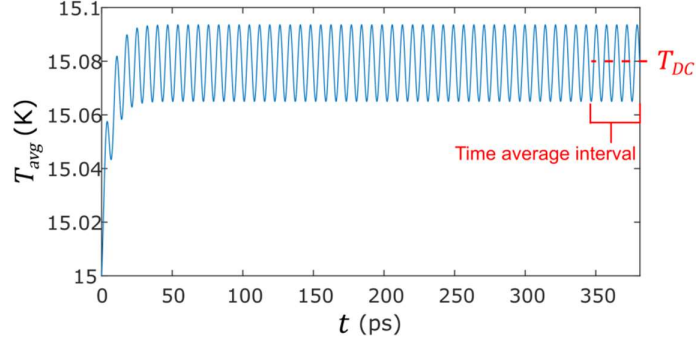

**Fig. S13** Simulated temporal oscillations in the spatial average of $T_e$ under oscillating heating power at $T_L$ = 15 K and a ratio $G_{WF}/G_{ep}$ = 70.3. After the heating begins, the oscillations reach a quasi-steady state. Averaging over the steady-state oscillations gives $T_{DC}$ = 15.079 K.

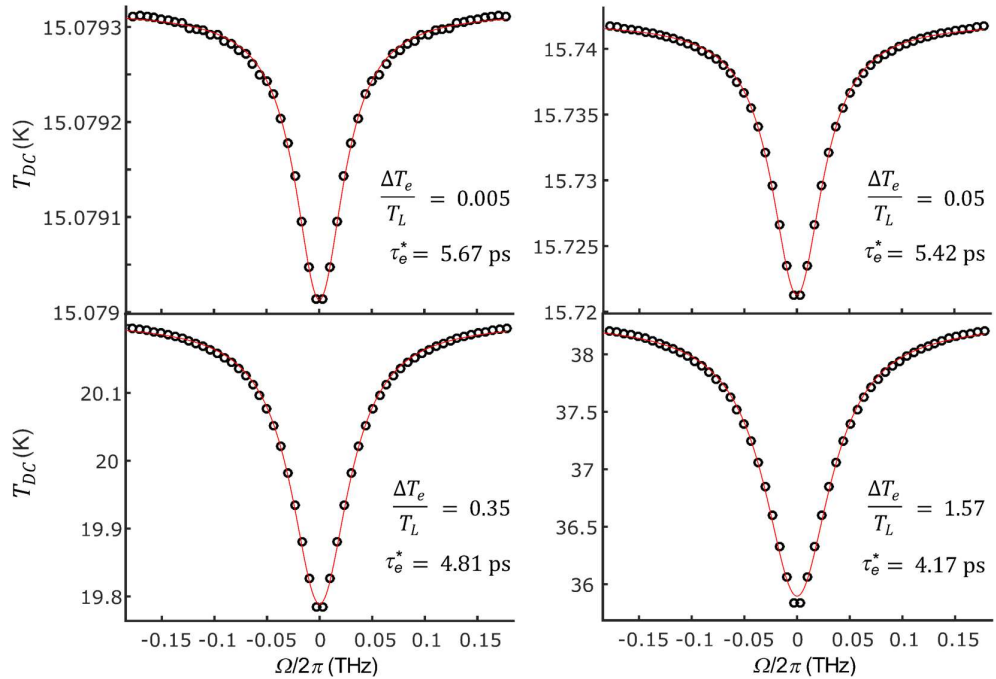

**Fig. S14** Calculated $T_{DC}$, from oscillations as in Fig. S13, as a function of the beating frequency for four different magnitudes of heating power at $T_L$ = 15 K and a ratio $G_{WF}/G_{ep}$ = 70.3 and a calculated $\tau_e$ = 6.90 ps. $\Delta T_e$ is estimated as $T_{DC} - T_L$ at the base of the Lorentzian. Curves are fit to a Lorentzian function to extract $\tau_e^*$. As heating power increases, the fractional uncertainty in the FWHM of these Lorentzians also increases, which may indicate that $T_{DC}$ vs $\Omega$ is not a strict Lorentzian function for finite heating power.

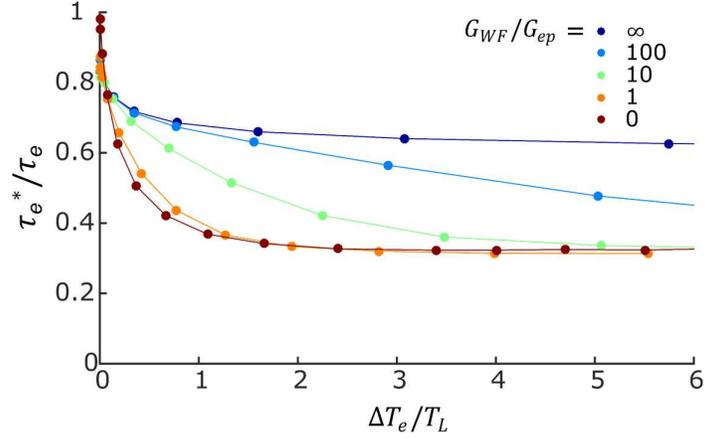

**Fig. S15** Calculated $\tau_e^*/\tau_e$ vs $\Delta T_e/T_L$ for five different ratios of $G_{WF}/G_{ep}$.

We investigated what factors affect the shape of the curve $\tau_e^*/\tau_e$ vs $\Delta T_e/T_L$ and found that there is fundamentally only one factor: the ratio between the strengths of cooling by heat diffusion and electron-phonon inelastic scattering. We quantify this ratio by the value $G_{WF}/G_{ep}$, where $G_{WF}$ and $G_{ep}$ are the thermal conductances of these two cooling mechanisms respectively. The curves are unaffected if the total cooling power is changed while maintaining the same ratio $G_{WF}/G_{ep}$, and they are also unaffected by any change in $C_e$. Note that the shape *is* affected by a change in aspect ratio of the graphene because the net strength of diffusive cooling is dependent on device geometry while the net electron-phonon cooling is only dependent on device area. We calculate $G_{WF}/G_{ep}$ from the expressions for each thermal conductance: $G_{WF} = \beta \mathcal{L}_0 T_L/R_{ch}$ and $G_{ep} = \delta A \Sigma T_L^{\delta-1}$. $\beta$ is a constant that depends on the aspect ratio of the device. Thus, we can tune $G_{WF}/G_{ep}$ in our simulation by changing $\beta$, $A$ or $\Sigma$. We then calculate $\tau_e$ as $C_e/(G_{WF} + G_{ep})$.

No matter what the value of $G_{WF}/G_{ep}$ is, the ratio $\tau_e^*/\tau_e$ approaches 1 as $\Delta T_e/T_L$ approaches zero. This is the same conclusion that we presented in the analysis of the previous section, "Derivation of the Lorentzian form of $\langle T_e(\Omega) \rangle$". These numerical simulations also demonstrate that $\tau_e^*/\tau_e$ decreases monotonically as a function of $\Delta T_e/T_L$. We next experimentally verify that this monotonic decrease does indeed occur. Figure S16 shows $\tau_e^*$ as a function of $\Delta T_e/T_L$ for two different densities and lattice temperatures.

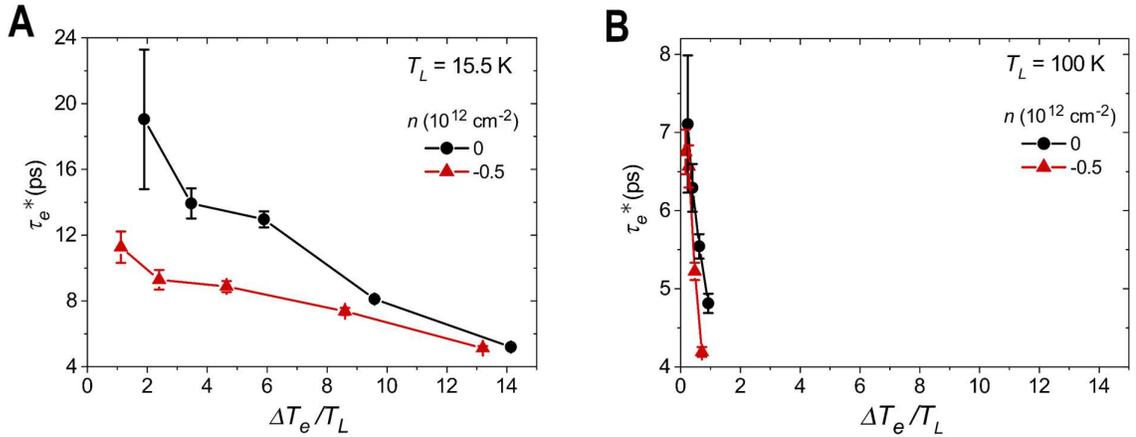

**Fig. S16** $\tau_e^*$ measured as function of $\Delta T_e/T_L$ at charge neutrality and for a density of $-0.5 \times 10^{12}$ cm$^{-2}$ when **(A)** $T_L = 15.5$ K and **(B)** $T_L = 100$ K.

We now seek to estimate the value of $\tau_e$ from our measurements of $\tau_e^*$ based on our understanding of the dependence of $\tau_e^*$ on the laser heating power. There are two approaches available to us. We may either 1) Extrapolate the measured $\tau_e^*$ vs $\Delta T_e/T_L$ to zero $\Delta T_e/T_L$, or 2) Divide $\tau_e^*$ by the ratio $\tau_e^*/\tau_e$ that we calculate from the simulation. Here, we have chosen to estimate $\tau_e$ by the latter approach because our fit of $G_{th}$ vs $T_e$ provides a fair estimate of the ratio $G_{WF}/G_{ep}$ that we need to perform this calculation. Also, the former approach has the disadvantage of requiring a time-consuming measurement of $\tau_e^*$ at several laser powers for each experimental condition.

We must calculate the curve $\tau_e^*/\tau_e$ vs $\Delta T_e/T_L$ corresponding to the experimental condition at which we have measured $\tau_e^*$ and then divide $\tau_e^*$ by the value of the curve at the $\Delta T_e/T_L$ used in the experiment. Figure S17 shows $\tau_e^*/\tau_e$ vs $\Delta T_e/T_L$ calculated for conditions corresponding to charge neutrality at several values of $T_L$. In these graphs we also plot a red interpolated data point at the value of $\Delta T_e/T_L$ used in the measurement of $\tau_e^*$ at the respective condition. Over the experimental range of $T_L$ (15-200 K), $G_{WF}/G_{ep}$ at charge neutrality ranges from 70.3 to 0.13, and we estimate $\tau_e^*/\tau_e$ to range from 0.5 to 0.7.

$G_{WF}/G_{ep}$ should also change with changing carrier density because both $G_{WF}$ and $G_{ep}$ are independent quantities which are known to increase with density (*4*). It would therefore be ideal to measure $G_{WF}/G_{ep}$ by a fit of $G_{th}$ vs $T_e$ over the whole investigated range of carrier densities. However, at most densities our $G_{th}$ vs $T_e$ data is too sparse at high temperatures to allow for sufficiently accurate fitting. Instead, we approximate $G_{WF}/G_{ep}$ to be constant with carrier density, and we estimate the uncertainty of this approximation from the range of possible values that take into account the theoretical proportionality of the electron-phonon coupling constant with carrier density and the inverse proportionality of $G_{WF}$ with $R_{ch}$. At the largest carrier density, our uncertainty in $G_{WF}/G_{ep}$ is greatest, with maximum and minimum possible values differing by a factor of 16 at each $T_L$. This typically results in a spread of 0.1 in the possible value of $\tau_e^*/\tau_e$, and this increases the uncertainty in the measurement of $\tau_e$ by adding to the total width of the error interval by 18% of the measured value of $\tau_e$ in the most extreme case. However, this error is always less than the error contributed by uncertainty in the Lorentzian peak fitting, and its contribution becomes smaller with decreasing carrier density.

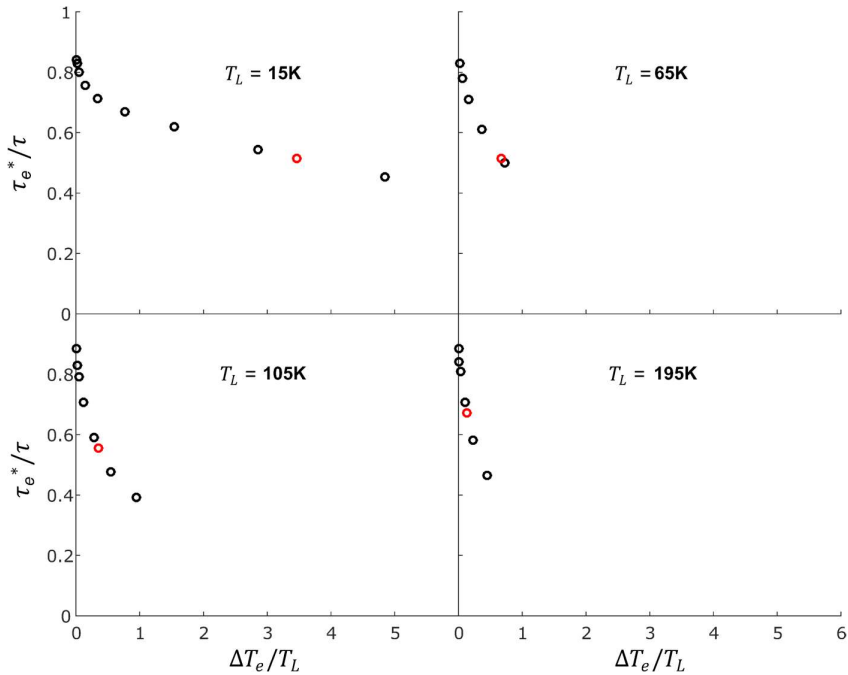

**Fig. S17** Black markers: $\tau_e^*/\tau_e$ vs $\Delta T_e/T_L$ calculated for values of $T_L = $ 15, 65, 105 and 195 K at charge neutrality. At these conditions $G_{WF}/G_{ep}=$ 70.3, 2.32, 0.71 and 0.13 respectively. Red markers: Interpolated points at the values of $\Delta T_e/T_L$ used during measurement of $\tau_e^*$.

## 8. Derivation of "$C_e = G_{th}\tau_e$"

An alternative derivation may be found in Ref. (5).

This equation relates the values of $C_e$, $G_{th}$ and $\tau_e$ at a specific temperature $T_L$. Consider a system temperature $T$ excited above $T_L$ by an amount $\Delta T$ so small that $C_e$, $G_{th}$ and $\tau_e$ do not change. That is to say

$$C_e, G_{th} \text{ and } \tau_e \text{ are constant for } T_L \leq T \leq T_L + \Delta T.$$

In this perturbative regime, after the exciting impulse is removed at time $t = 0$, $T$ will relax back to $T_L$ exponentially:

$$T(t) = \Delta T e^{-t/\tau_e} + T_L$$

Also, as this relaxation progresses, the power that cools the system will diminish as

$$\dot{Q}_{cool} = G_{th}[T(t) - T_L]$$

The differential equation dictating the time evolution of $T$ is

$$C_e \frac{dT(t)}{dt} = -\dot{Q}_{cool}$$

which simplifies as follows by integrating over the relaxation process:

$$C_e \int_{T_L+\Delta T}^{T_L} \frac{dT(t)}{dt} dt = -G_{th} \int_0^\infty [T(t) - T_L] dt$$

$$C_e = G_{th}\tau_e$$

## 9. Extended Data

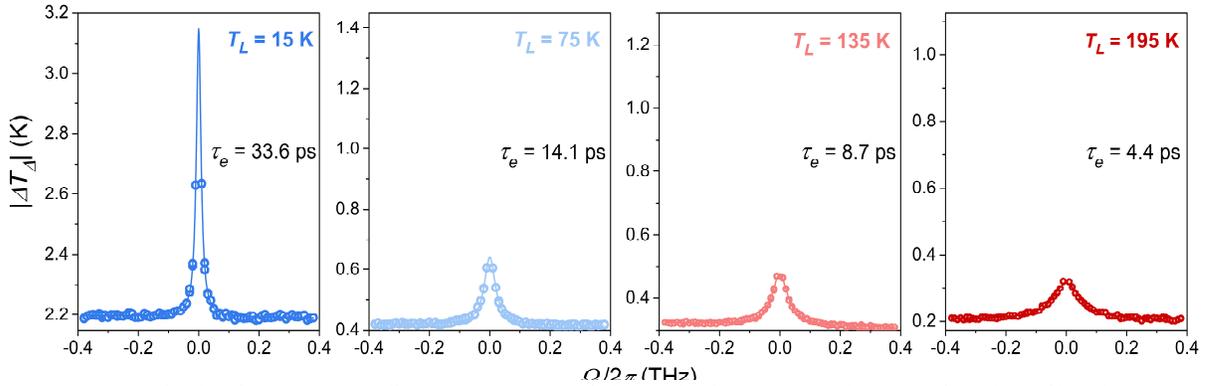

**Fig. S18** Magnitude of $\Delta T_\Delta$, the non-linear component of change in time-averaged $T_e$, as a function of beating frequency $\Omega$ at several values of $T_L$ at charge neutrality; solid lines are Lorentzian function fits of width $(\pi\tau_e)^{-1}$, with the normalized values of $\tau_e$ stated.

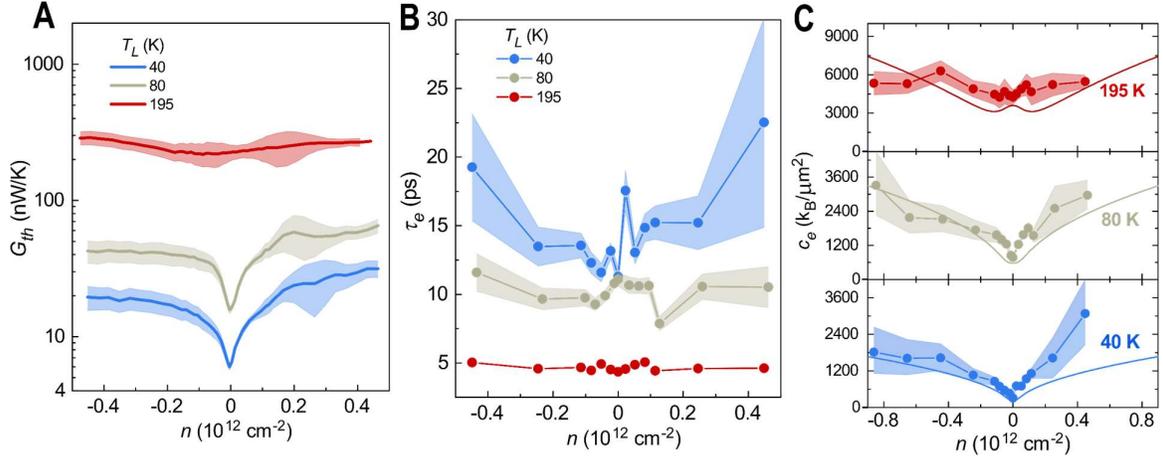

**Fig. S19** Experimental results for **(A)** thermal conductance, **(B)** thermal relaxation time and **(C)** heat capacity per unit area measured in the same device used in the main text at three different lattice temperatures $T_L$.

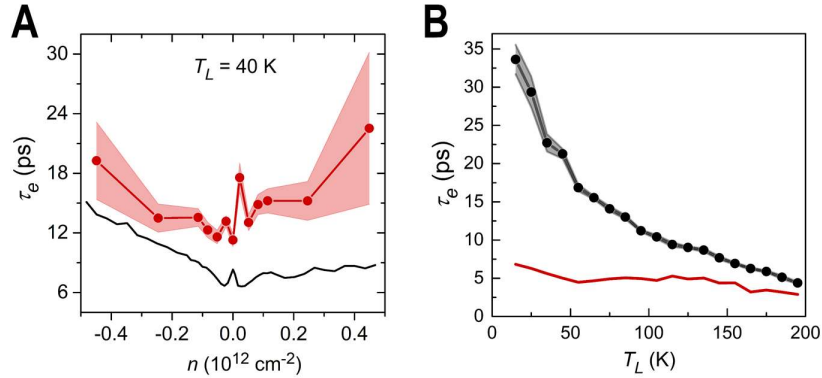

**Fig. S20 (A)** (Black) Measured thermal relaxation time vs carrier density at 40 K lattice temperature and (Red) calculated thermal relaxation time at 40 K determined from the measured thermal conductance and the theoretical heat capacity according to $\tau_e^{calc} = C_e^{theory}/G_{th}^{expt}$. **(B)** (Black) Measured thermal relaxation time vs temperature at charge neutrality and (Red) calculated thermal relaxation time at charge neutrality according to $\tau_e^{calc} = C_e^{theory}/G_{th}^{expt}$.

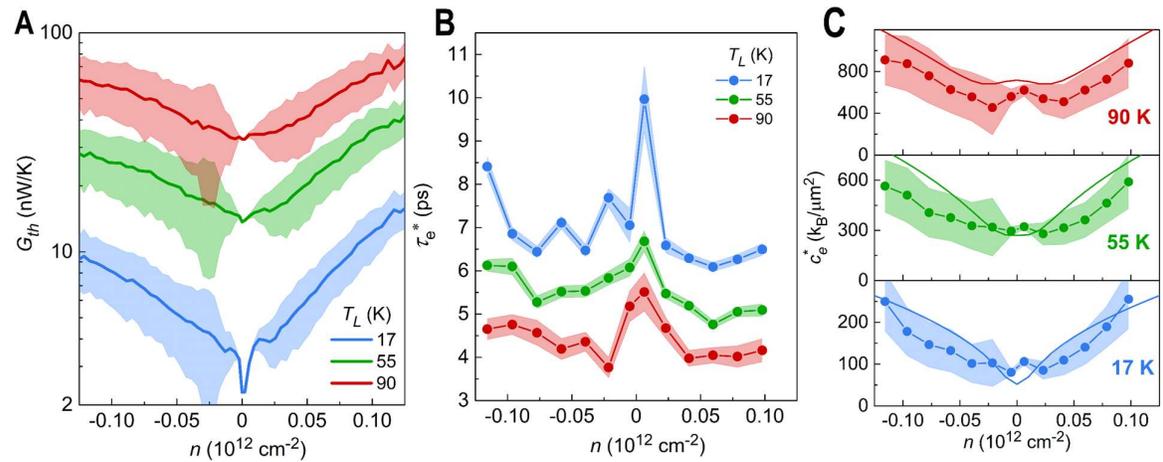

**Fig. S21** Experimental results obtained in another monolayer two-terminal graphene device originating from a another exfoliated flake. Plots are of **(A)** thermal conductance, **(B)** the un-normalized thermal relaxation time and **(C)** the heat capacity estimate based on product of these two. Thermal relaxation time cannot be normalized because $G_{th}$ data is too sparse to estimate the ratio $G_{WF}/G_{ep}$ needed for the calculation.

# 10. Theoretical Calculation of $C_e$ of Graphene charge carriers

Electronic heat capacity $c_e$ can be calculated from

$$c_e = \int_0^\infty dE\,(E - E_F)\frac{df}{dT_e}g(E)$$

where $E$ is the energy of the energy levels occupied by the charge carriers obeying the Fermi-Dirac distribution given by

$$f(E) = \frac{1}{e^{(E-\mu)/k_B T_e} + 1}$$

with $\mu$ the chemical potential which is equal to $E_F$ at $T_e = 0\,K$. The density of states is expressed by $g(E) = (2/\pi\hbar^2 v_f^2)E = \gamma E$. Plugging in the above expressions yields

$$c_e = k_B\left[\gamma(k_B T_e)^2\int_0^\infty \frac{e^x x^3}{(e^x + 1)^2}dx + k_B T_e \gamma\mu\int_0^\infty \frac{e^x x^2}{(e^x + 1)^2}dx\right]$$

$$c_e = \left[5.409(k_B T_e)^2 + \left(\frac{\pi^2}{3}\right)\mu k_B T_e\right]\gamma k_B$$

in the approximation $\mu - E_F \approx 0$ which is not exact at all conditions (see Fig. S21). The first term is a quadratic dependence on $T_e$ which is a unique feature to Dirac electrons as a consequence of their linear dispersion. It is dominant over the second term in the undoped case where $\mu$ is zero or extremely small. The second term is the conventional linear-in-$T_e$ contribution that occurs in all Fermi liquids (of all dimensions).

In order to accurately calculate $\mu$, we relate it with the carrier density $n$ by enumerating all the carriers occupying energy levels (labelled by wavevectors $\vec{k}$) of the linear bands with dispersion given by $E = \hbar v_f|\vec{k}|$:

$$n = \frac{4}{A}\sum_{\vec{k}}[f(E) + (1 - f(-E))]$$

where a factor of 4 accounts for the spin and valley degeneracy and $A$ is the area of the graphene sheet. The summation, performed over all $\vec{k}$ for all electrons in both the conduction band and valence band, is converted to the integral $\left(\frac{A}{2\pi}\right)\int_0^\infty k\,dk$, where $k = |\vec{k}|$, leading to

$$n = \frac{2}{\pi}\int_0^\infty dk\,k\left[\frac{1}{e^{(\hbar v_f k - \mu)/k_B T_e} + 1} - \frac{1}{e^{(\hbar v_f k + \mu)/k_B T_e} + 1}\right]$$

$$\left(\frac{E_F}{k_B T_e}\right)^2 = 2\int_0^\infty dx\,x\left[\frac{1}{e^{x - \mu/k_B T_e} + 1} - \frac{1}{e^{x + \mu/k_B T_e} + 1}\right]$$

where we have used the relationship $E_F = \hbar v_f\sqrt{\pi n}$. The resultant dependence of $\mu$ on $E_F$ is plotted in Fig. S22.

Extending it further to calculate the total energy $U$ by summing the energy $E$ over all energy levels:

$$\frac{U}{A} = \frac{4}{A}\sum_{\vec{k}}\{E[f_T(E) - \theta(E_F - E)] + (-E)[f_T(-E) - \theta(E_F + E)]\}$$

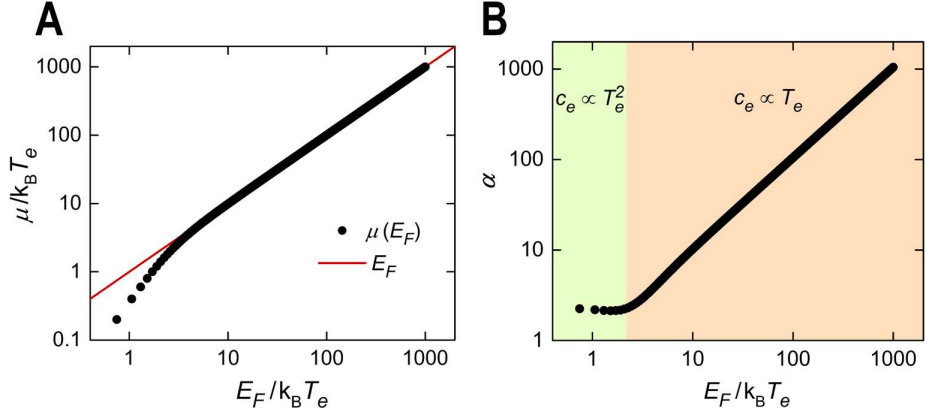

**Fig. S22 (A)** Chemical potential as a function of Fermi energy. **(B)** $\alpha$ as a function of Fermi energy.

The summation is performed over all $\vec{k}$ for all electrons in the conduction band with energy $E$ and for holes in the valence band with energy $-E$, as encoded in the summand above. $\theta(E_F - E)$ is the step function representing the energy level occupancy at $T_e = 0$ K for the conduction band, which is subtracted as zero thermal energy reference whereas $\theta(E_F + E)$ is the valence band counterpart.

Similar evaluation of $U/A$ leads to

$$\frac{U}{A} = \frac{2}{\pi} \int_0^\infty dk\, k^2 \left[ \frac{1}{e^{(\hbar v_f k - \mu)/k_B T_e} + 1} + \frac{1}{e^{(\hbar v_f k + \mu)/k_B T_e} + 1} - 1 - \theta(E_F - \hbar v_f k) + \theta(E_F + \hbar v_f k) \right]$$

Further evaluation leads to

$$\frac{U}{A} = \frac{(k_B T_e)^3}{(\hbar v_f)^2} \alpha$$

where

$$\alpha = \frac{2}{\pi} \int_0^\infty dx\, x^2 \left[ \frac{1}{e^{x - \mu/k_B T_e} + 1} + \frac{1}{e^{x + \mu/k_B T_e} + 1} \right] - \frac{2}{3\pi} \left( \frac{E_F}{k_B T_e} \right)^3$$

$\alpha$ is effectively only a function of $E_F/k_B T_e$ which is plotted in Fig. S22.

The electronic heat capacity is finally derived by the derivative:

$$c_e = \frac{d}{dT_e}\left(\frac{U}{A}\right) = \frac{d}{dT_e}\left[\frac{(k_B T_e)^3}{(\hbar v_f)^2} \alpha\right]$$

Following the numerical computations, $c_e$ is plotted as a function of $T_e$ and $n$ in Fig. S23.

For sufficiently low $E_F/k_B T_e \ll 1$, where $\alpha$ is a constant as highlighted in Fig. S22, we find $c_e(T_e) \propto T_e^2$ as expected, whereas at higher $E_F/k_B T_e > 1$, we find $c_e(T_e) \propto T_e$.

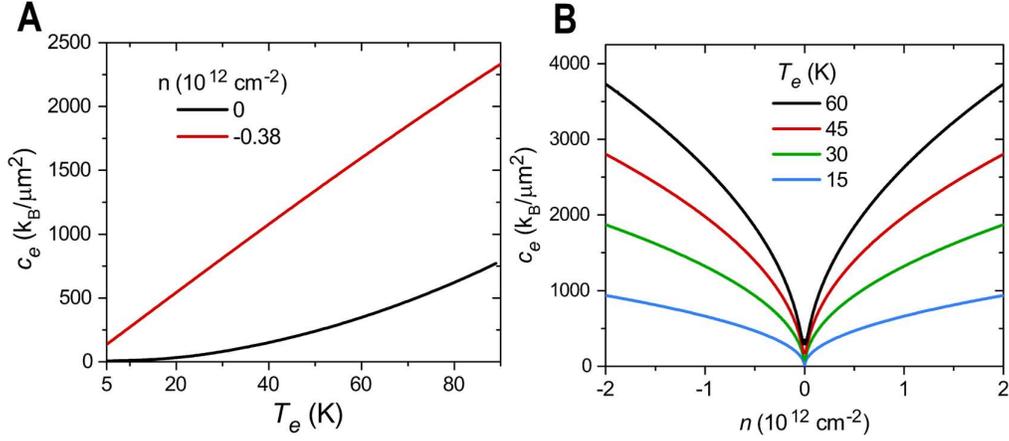

**Fig. S23** Numerically calculated Electronic heat capacity of graphene as a function of **(A)** temperature and **(B)** carrier density.

In order to account for spatial carrier density fluctuations which are pronounced at the charge neutrality point, we use a Gaussian probability distribution of carrier density with $n$ as the mean carrier density, $n'$ as the local carrier density and $\sigma_n$ as the standard deviation

$$P(n';n) = \frac{1}{\sigma_n\sqrt{2\pi}} e^{\frac{1}{2}\left(\frac{n'-n}{\sigma_n}\right)^2}$$

The electronic heat capacity of graphene because of this distribution of carrier density is

$$\bar{c}_e(n) = \int c(n') P_n(n';n) dn'$$

$$= \frac{1}{\sigma_n\sqrt{2\pi}} \int c(n') e^{\frac{1}{2}\left(\frac{n'-n}{\sigma_n}\right)^2} dn'$$

We calculate $\bar{c}_e(n)$ in a MATLAB routine as a convolution of $c_e(n)$ with the exponential $e^{\frac{1}{2}\left(\frac{n}{\sigma_n}\right)^2}$ and show the results in Fig. S24. We estimate $\sigma_n$ from the $1/R$ vs $n$ plot as shown in Fig. S10.

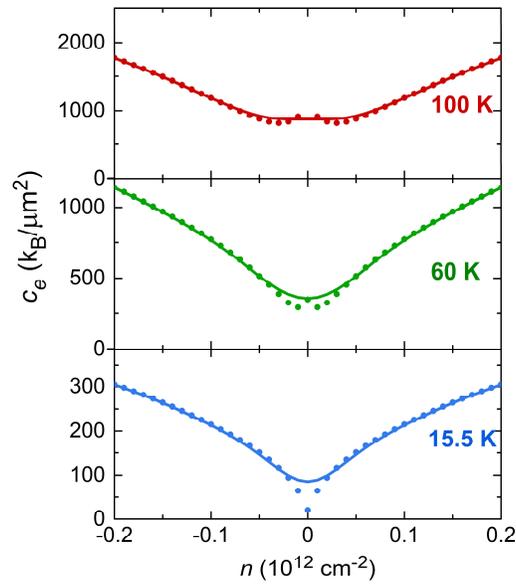

**Fig. S24** Calculated electronic heat capacity per unit area with spatial carrier density fluctuations accounted for (solid lines) and with fluctuations neglected (circles).

1. R. H. Dicke, The Measurement of Thermal Radiation at Microwave Frequencies, Rev. Sci. Instrum. 17, 268 (1946).
2. Xia, F., Perebeinos, V., Lin, Y. M., Wu, Y. & Avouris, P. The origins and limits of metal-graphene junction resistance. *Nature Nanotechnology* **6**, 179–184 (2011).
3. Crossno, J. *et al.* Observation of the Dirac fluid and the breakdown of the Wiedemann-Franz law in graphene. *Science* **351**, 1058–1061 (2016).
4. Fong, K. C. *et al.* Measurement of the Electronic Thermal Conductance Channels and Heat Capacity of Graphene at Low Temperature. *Physical Review X* **3**, 041008 (2013).
5. Bachmann, R. *et al.* Heat capacity measurements on small samples at low temperatures. *Review of Scientific Instruments* **43**, 205–214 (1972).